      [1999/12/01 v1.4c Il Nuovo Cimento]
\documentclass[varenna]{cimento}

\usepackage{graphicx}  
\title{Galaxy systems in the optical and infrared}

\author{Andrea Biviano}

\institute{INAF/Osservatorio Astronomico di Trieste \\ 
via G.B. Tiepolo 11, 34143 -- Trieste, Italy}

\shortauthor{A. Biviano}
\shorttitle{Galaxy systems in the optical and infrared}

\begin{document}

\maketitle

\begin{abstract}
  In these three lectures a review is provided of the properties of
  galaxy systems as determined from optical and infrared measurements.
  Covered topics are: clusters identification, global cluster
  properties and their scaling relations, cluster internal structure
  and dynamics, and properties of cluster galaxy populations.
\end{abstract}

\section{Identification, global properties, and scaling relations}
\subsection{Identification}
\label{ident}
Historical identifications of clusters of {\em nebul\ae}~ date back to
the late years of the XVIII century~\cite{bivi00}. The first modern
method of galaxy clusters identification and classification was
implemented by Abell~\cite{abel58} in 1958. Abell worked out apparent
overdensities of galaxies in the sky by eye inspection of photographic
plates of the Palomar Observatory Sky Survey.  He used the apparent
magnitudes of galaxies to determine approximate cluster distances
which he then used to convert apparent sizes to physical sizes. Abell
characterized clusters by their richness, i.e. the number of galaxies
in the 2 mag range $m_3$ to $m_3+2$\footnote{$m_3$ is the magnitude of
the third brightest galaxy in the cluster field.} in a circle of
radius 2.1 Mpc\footnote{$H_0=70$ km~s$^{-1}$~Mpc$^{-1}$,
$\Omega_m=0.3$, $\Omega_{\Lambda}=0.7$ are adopted throughout these
lectures.}, corrected for the contamination by fore- and back-ground
galaxies. We now refer to this radius as the 'Abell radius'; it is
still widely used as a typical cluster size since it is rather close
to the virial radius, $r_{200}$\footnote{The virial radius $r_{200}$
is the radius within which the enclosed average mass density of a
cluster is 200 times the critical density, $200 \, \rho_c$.}, of a
massive cluster with a velocity dispersion $\sigma_v \simeq 900$
km~s$^{-1}$.
 
Abell's catalog contained 2712 clusters, later extended to the
southern hemisphere~\cite{abel89} to a total of 4073 clusters. Thanks
to Abell's early work, galaxy clusters started to be studied as a
class and not only as individual objects. Abell's catalog has formed
the basis of the largest galaxy cluster-specific spectroscopic survey
completed so far, the ESO Nearby Abell Cluster Survey
(ENACS \cite{katg96,katg98}).

Abell's cluster sample suffers however from two main problems,
incompleteness and contamination by projection
effects~\cite{luce83,katg96}, although these problems are less severe
than it is usually stated~\cite{brie93}.  Incompleteness and
contamination are crucial issues, in particular for cosmological
studies, like e.g. the determination of the cluster mass function and
its redshift evolution (see Sect.~\ref{cosmo}). Ideally one would like
to have a cluster sample with zero contamination and 100\%
completeness. Since this is not possible, it becomes essential that
contamination and incompleteness can be precisely estimated, so that
the statistical sample can be corrected for. A precise estimate of the
contamination and incompleteness of a cluster catalog can be obtained
by applying the cluster identification technique to mock galaxy
samples extracted from cosmological numerical simulations. This
of course requires the identification technique to be exactly
reproducible, and Abell's eye is not.

The main difference of today's clusters identification techniques with
respect to Abell's is that they are both automated and objective.
Automatization requires either that photographic plates be digitized,
or that data are digital in origin, coming from CCD cameras
\cite{gal03}. Besides automatization and objectiveness, modern
clusters identification techniques also have a well-understood
selection function and impose minimal {\em a priori} constraints on
the properties of the systems to be identified
\cite{koes07}. These characteristics are now common to many
cluster identification methods which have been developed over the
years. Among these, some have been specifically developed to be used
on photometric galaxy samples and some on spectroscopic galaxy
samples.

Among the methods applicable to samples of galaxies without redshift
information, the most used is the Matched Filter (MF
hereafter \cite{post96}).  In summary, the method works as follows.
The spatial and luminosity distribution of observed galaxies in a
given field is modeled as the sum of two contributions, one from the
field, another from the cluster
\begin{equation}
D(R,m)=b(m)+ N_c N(R) \phi(m).
\end{equation}
The term $b(m)$ represents the background galaxy
counts at a given magnitude, $m$. The cluster term is itself the
product of three factors. $N(R)$ is the projected radial profile of
the cluster galaxies as a function of the projected radial distance
from the cluster center, $R$, $\phi(m)$ is the differential cluster
luminosity function (LF hereafter; see Sect.~\ref{glob}), and
$N_c$ is a measure of the cluster richness, or multiplicity.
Both $N(R)$ and $\phi(m)$ can depend on free parameters, typically a
characteristic length scale and a characteristic magnitude.  The
best-fit free parameters are found through a Maximum Likelihood
procedure aimed at minimizing the difference between the observed
galaxy distribution, $D(R,m)$, and the model. Clusters are identified
by searching for local maxima within a moving box of given size
centered on each pixel of the filtered galaxy map array (or at each
galaxy position~\cite{kepn99}).  If the central pixel in the box is a
local maximum, and if the maximum exceeds a given threshold (which
depends on the background noise), a candidate cluster is
registered. An estimate of the cluster redshift results from assuming
a universal value for the absolute characteristic magnitude of the
LF.  Several variants of the MF method have
been proposed~\cite{kepn99,lobo00,dong08}.

Among the methods applicable to samples of galaxies with redshift
information, by far the most used is the friends-of-friends
percolation algorithm (FoF hereafter, see
\cite{huch82,gell83}). This method links together all galaxies within a
chosen linking volume centered on each galaxy. At variance with
Abell's method, galaxy systems are identified within a physical
overdensity, not within a physical (fixed) size. Since the density of
galaxies in a flux-limited survey depends on redshift, the linking
volume is also scaled with redshift. In practice, specifying the
linking volume is equivalent to specifying two linking lengths, one in
the plane of the sky, another along the redshift direction. Different
works have adopted different linking lengths, and different scalings
with the galaxy density (compare, e.g.,
\cite{huch82} to \cite{rame89}). The linking lengths have been chosen
using {\em a priori} knowledge of the physical characteristics of the
galaxy systems one is looking for \cite{eke04}, or by minimizing the
differences between the recovered and intrinsic properties of systems
identified in a mock galaxy sample \cite{berl06}.

While originally conceived to work on imaging surveys, in its modified
versions the MF method can also be applied to spectroscopic surveys
\cite{kepn99,dong08}. Symmetrically, when the
redshift information is not available, it is still possible to adopt
the FoF method, using photometric, rather than spectroscopic,
redshifts \cite{vbre07}.  Nevertheless, applications of the MF
(respectively, FoF) method have so far mostly concerned data from
photometric (respectively, spectroscopic) surveys.

The MF algorithm has been applied to data from the ESO Imaging
Survey \cite{olse99,lobo00}, the Sloan Digital Sky Survey
(SDSS \cite{bahc03}), the 2 Micron All Sky Survey
(2MASS \cite{koch03}), and several other surveys
\cite{will01,dona02,post02,diet07,vbre07}.  Clusters in the resulting
catalogs are detected out to $z \sim 1$ and beyond, and down to masses
$\sim 10^{14} \, M_{\odot}$, depending on the depth of the photometric
survey. Spectroscopic follow-ups show the photometrically identified
high-$z$ cluster candidates to be real \cite{beno02,olse05}.
Comparison to mock catalogs show that completeness can reach $\sim
100$\% for $\sim 10^{14} \, M_{\odot}$ mass clusters out to
intermediate-$z$, and $\sim 50$\% for very massive clusters ($\sim
10^{15} \, M_{\odot}$) out to $z \sim 1.5$ \cite{vbre07,dong08}. In
comparison to X-ray cluster surveys, these optical cluster surveys
have been able to detect lower-mass clusters \cite{dona02,basi04} and
clusters with an X-ray luminosity below what expected given their mass
(see, e.g., \cite{bowe97,mcna01,pope07}).  Less than 10\% of X-ray
detected clusters with $z \leq 0.5$ are missed in these optical cluster
surveys \cite{schu04}.

The FoF method has been applied to many spectroscopic
surveys, e.g. the Center for Astrophysics Redshift Survey
\cite{gell83,rame89,rame97}, the Southern Sky Redshift Survey
\cite{maia89}, the Las Campanas Redshift Survey \cite{tuck00}, the ESO
Slice Project survey \cite{rame99}, and, more recently, the SDSS, the
Two Degree Field Galaxy Redshift Survey (2dFGRS), and the 2-Micron All
Sky Redshift Survey (2MRS)
\cite{eke04,merc05,berl06,croo07,deng07,tago08}.  The resulting
catalogs list up to several thousands of galaxy systems.  Most of them
are small galaxy groups, with median velocity dispersion and mass $\sigma_v
\sim 200$ km~s$^{-1}$, $M \sim 10^{13} \, M_{\odot}$.  
Clearly, identification in 3-d space (spatial coordinates and
redshifts) allows to detect lower-mass systems than identification in
projected spatial distribution only. On the other hand, the higher depth of
photometric surveys allows the detection of (rather massive) clusters
out to higher-$z$.

Several other cluster identification methods have been implemented.
Some methods start with a FoF identification, then provide a first
estimate of the mass of the identified system, and use this estimate to
determine which galaxies belong to the group, in an iterative way
\cite{carl01,yang05,wein06}. These variants of the FoF method
try to minimize contamination by galaxies that are close in redshift
but not in real space (interlopers). This is crucial when the
contamination risk is high, like in medium- or high-$z$ samples. An
application of this technique to the Canadian Network for Observational
Cosmology (CNOC) $2^{nd}$ survey has produced a
catalog of $\sim 200$ galaxy groups at a median $z=0.33$ \cite{carl01}.

\begin{figure}
\center{\includegraphics[height=10cm]{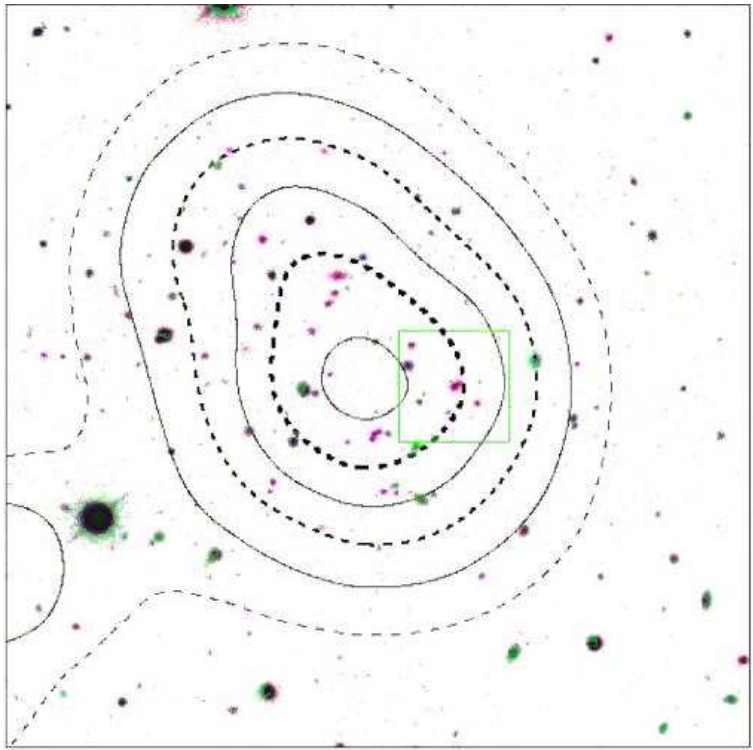}}
\caption{Color composite image of a cluster identified with the
  CRS technique \cite{glad05}. The cluster is located at an
  estimated redshift of 0.952. Red-sequence galaxies are displayed
  with a reddish color in the figure, and red-sequence galaxy
  isodensity contours are displayed. The rectangle identify the
  BCG. The image size is $15' \times 15'$,
  i.e. $\sim 7 \times 7$ Mpc at the cluster estimated distance. Note
  how the galaxy colors are helpful in identifying the cluster against
  the background.}
\label{fig-crs}
\end{figure}

The Cluster Red Sequence (CRS hereafter \cite{glad00}) method is based
on the observation that all rich clusters, at all redshifts up to
$z \sim 1$, have a more or less well defined red sequence of galaxies
in a color--magnitude diagram, where the color is defined by two
photometric bands bracketing the 4000~\AA~ break feature of galaxy
spectra. Since the 4000~\AA~ break is redshifted into different
observational bands depending on the galaxy $z$, with a suitable set
of filters it is possible to define color-cuts to select galaxies at a
redshift close to the cluster mean redshift. By comparison with
spectrophotometric models, the CRS method also provides estimates of
the mean redshifts of the detected clusters, with an accuracy superior
to that reached by the MF method \cite{glad00}. Roughly speaking, the
method consists in slicing a given galaxy catalog in color, computing
the galaxy surface density of that slice, and identifying significant
overdensities. An example of a cluster detected with this
technique \cite{glad05} at an estimated redshift $z \sim 1$ is shown
in Fig.~\ref{fig-crs}.

The CRS method has been first tested on the CNOC-2 data-set, for which
spectroscopy is available, using only two-band photometry.  In
practice, this is a comparison of the photometric CRS vs.  the
spectroscopic FoF methods. The fraction of clusters detected with the
CRS and undetected with the FoF method, i.e. the fraction of false
positive CRS detections, is only 1/23 out to $z=0.5$, and the CRS
photometric $z$-estimates are accurate to $\Delta z=0.03$. Given that
many of the identified systems are groups rather than rich clusters,
the performance of the CRS method is remarkable.  Application of this
method to mock data-sets extracted from the Millennium simulation has
shown that projection effects are of relatively minor importance in
the great majority of the identified systems (80 to 90\%, depending on
$z$) \cite{cohn07}.

The CRS method has been applied to photometric data obtained in two
wavebands with large format mosaic cameras at CFHT and CTIO. The
resulting catalog (the Red-sequence Cluster Survey, see
\cite{glad05,glad07}) contains almost 1000 cluster candidates 
among which more than a hundred at $z \sim 1$, some of which have been 
spectroscopically confirmed \cite{barr04,glad05}.

The CRS method has been applied to infrared (IR hereafter)
data obtained with the IRAC camera onboard {\em Spitzer}
\cite{eise08,wils08}. At higher $z$ the 4000~\AA~ break is progressively 
shifted towards the IR, hence deep IR observations are needed to
detect cluster red-sequences at $z>1$. Hundreds of candidate clusters
out to an estimated redshift $z=1.85$ have been found so far by this
method using data from Spitzer surveys, of which more than one hundred
above $z=1$ \cite{eise08,wils08}. Other spectral features than the
4000~\AA~ break can be used to define IR color cuts aimed at reducing
the field galaxy contamination, e.g.  the $\simeq 1.6 \, \mu$m peak in
the flux distribution of stellar populations \cite{papo08}.  Based on
this spectral feature $z>1$ clusters can be selected by using two {\em
Spitzer} IRAC bands (3.6 and 4.5 $\mu$m) which will still be in
operation in the {\em Spitzer} postcryogenic era.  Recently, data from
the Japanese IR satellite {\em AKARI} have been used to detect
candidate clusters at $0.9<z<1.7$ by exploiting both the 4000~\AA~
break and the $1.6 \, \mu$m peak features \cite{goto08}.

Another cluster identification method that relies upon the existence
of a red sequence for cluster galaxies, is maxBCG
\cite{bahc03,koes07}. The maxBCG method also relies on the location
of the brigthtest cluster galaxy (BCG) near the cluster center.
Application of this method to the SDSS data \cite{koes07b} has
provided a catalog of 13,823 clusters with $\sigma_v > 400$
km~s$^{-1}$, out to $z=0.3$, with photometric $z$ estimates accurate
to $\Delta z \simeq 0.01$.  Comparison with a cluster catalog
constructed by applying the MF method on the same data-set has shown that
$\sim 80$\% of the systems are identified by both the maxBCG and the MF
method \cite{bahc03}.  The imperfect matching may be due to the
presence of substructures (see Sect.~\ref{subs})
identified as distinct clusters by the
maxBCG method, and to the presence of false positives in both catalogs.
Comparison with mock catalogs indicates the maxBCG cluster catalog is
90\% pure and 85\% complete for clusters with masses $\geq 10^{14} \,
M_{\odot}$.

Relying upon the existence of a red sequence for cluster galaxies has
certainly proven to be a very effective way of selecting galaxy
clusters. On the other hand, unrelaxed, low-mass galaxy clusters in
which this sequence is not established yet, or at least not very
prominent \cite{dona02}, may be missed by CRS methods and alike. This
is particularly true at high $z$, since the fraction of early-type
galaxies (ETGs in the following) decreases with $z$ (see
Sect.~\ref{morph}) making the red sequence less
and less prominent.  For this reason, other cluster detection methods
make use of multi-band photometry to allow a wider selection of galaxy
spectral types, star-forming galaxies included. Typically, but not
exclusively, this is done by defining photometric-$z$ through the
comparison of the galaxy spectral energy distributions with model
templates. The Cut and Enhance (CE)
\cite{goto02} and the C4 methods \cite{mill05} 
have been developed to make full use of the SDSS 4-bands
photometry.

At high $z$, IR photometry proves essential to identify galaxy cluster
candidates. A clear demonstration of the potential of cluster searches
conducted in the IR is the detection of a $z=1.41$
cluster \cite{stan05}.  This cluster has been identified in the {\em
Spitzer} IRAC Shallow survey as an overdensity of IR-selected galaxies
(see also \cite{elst06,eise08}). Each galaxy was
assigned a photometric-redshift probability distribution, which was
then used to weigh the density maps within overlapping redshift
slices. A wavelet technique was adopted to smooth the density field,
in which significant peaks were then looked for.  The $z=1.41$ cluster
is only the highest-$z$ confirmed detection among 335 galaxy cluster
and group candidates (average mass $\sim 10^{14} \, M_{\odot}$) found 
with this technique in a 7.25 deg$^2$ region in the {\em Spitzer} 
IRAC Shallow survey \cite{brod08,eise08}. 
Over a hundred of the cluster candidates have a redshift above unity,
with an estimated spurious detection rate of 10\%.  Twelve of the
$z>1$ candidates have already been spectroscopically
confirmed. Unfortunately, the photometry is not deep enough to
identify clusters much beyond $z \approx 1.5$ \cite{brod08}.
 
IR-color selection is also useful to improve the efficiency of
spectroscopic follow-ups of high-$z$ clusters identified in other
wavebands, e.g. in X-rays \cite{gill03,kurk08}. 

\begin{figure}
\center{\includegraphics[height=9cm]{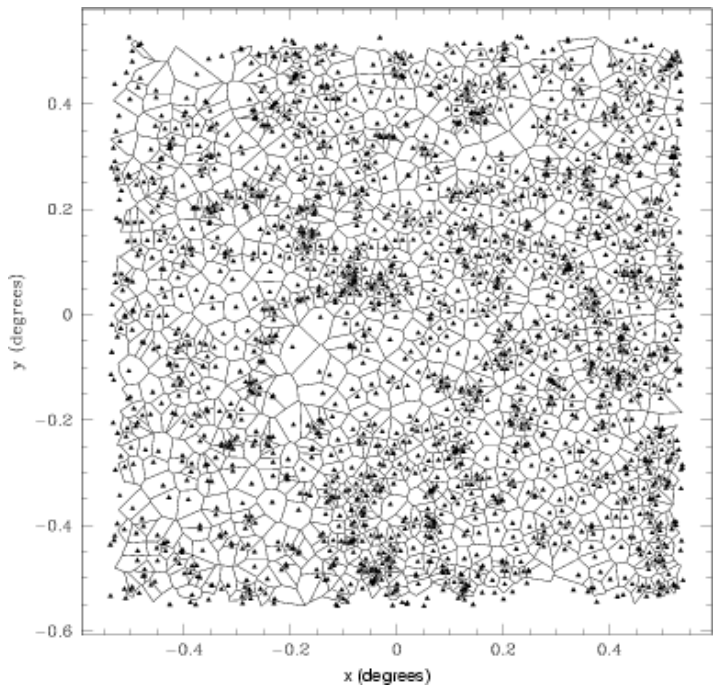}}
\caption{An example of Voronoi tesselation of a galaxy field. Each
  triangle represents a galaxy position on the plane of the sky 
  \cite{rame01}.}
\label{fig-voro}
\end{figure}

A rather different method for the identification of galaxy systems is
the Voronoi Galaxy Cluster Finder (VGCF \cite{rame01}).
In this method, the projected space is divided in cells according to
the Voronoi tesselation technique, each cell containing a single point
(i.e. a galaxy; see Fig.~\ref{fig-voro}).  The inverse of the cell area
defines the local galaxy density.  Clusters are defined as ensembles
of adjacent cells with a density above a given threshold. The search
for clusters is done in magnitude bins. The main advantages of this
method is that it is nonparametric and as such it does not require
{\em a priori} hypotheses on the cluster properties, such as cluster
size, density profile, or shape. The VGCF method has been
shown to be competitive with (or even better than) the MF
method \cite{barr05}.  A comparison between
the CE, MF, maxBCG, and
VCGF methods has shown that at low redshifts ($z < 0.4$) the
CE is more complete, but pays the price of a
higher rate of false detections ($\sim 30$\%) as measured from Monte
Carlo simulations \cite{goto02}.

If multicolor photometry is available, the VGCF method can be
applied to subsamples selected in the color-magnitude diagram, thus
reducing the field contamination by the same technique adopted in
the CRS method \cite{kim02}. This VGCF+CRS
method has been used to detect clusters in fields previously observed
by the {\em Chandra} X-ray satellite \cite{bark06}. The optical
detection fraction of X-ray-detected clusters was 46\% vs. an X-ray
detection fraction of optically-detected clusters of only 11\%. While
part of the optical detections may be spurious, the cluster detection
threshold was clearly lower in the optical catalogs than in the X-ray
catalogs. Galaxy-rich clusters without X-ray counterparts are also
detected, suggesting the existence of a population of underluminous
(perhaps not yet virialized) X-ray clusters (see,
e.g., \cite{bowe97,mcna01,pope07}).  Sufficiently deep photometric
data-sets allow the VGCF method to detect clusters as far as
$z \sim 1$ and beyond \cite{vbre07}.

With some modifications \cite{mari02}, the VGCF method has
also been used on 3-d data-sets, such as the DEEP2 Galaxy Redshift
Survey, resulting in a catalog of 105 galaxy systems with median
$z=0.86$, and median $\sigma_v=480$ km~s$^{-1}$ \cite{gerk05}.

Eventually, when the data are too shallow to rely on galaxy number
counts, it is possible to identify cluster candidates as positive
surface-brightness fluctuations (SBF) in the background
sky \cite{gonz01,bart02}. An application of the SBF method to a
drift-scan survey has produced a catalog of $\sim 1000$ cluster
candidates at $z \leq 0.8$, the Las Campanas Distant Cluster Survey
(LCDCS \cite{gonz01,gonz02}). At $z<0.3$, the catalog contains systems
with masses typical of galaxy groups. This catalog has formed the
basis for the ESO Distant Cluster Survey (EDisCS \cite{whit05}), in
which 20 LCDCS candidate clusters have been followed up
spectroscopically with VLT/FORS2.

Going to the next step of the cosmic hierarchy, catalogs of clusters
can be used to define superclusters \cite{abel61}. Superclusters have
been identified either by the FoF percolation technique
\cite{bahc84,batu85,zucc93,eina97}, or as overdensities in smoothed
cluster-density fields \cite{kali95,eina06,eina07}. Large-scale
structure morphology has also been characterized statistically,
without the need of defining superclusters, e.g. by the Genus
statistics \cite{gott86}, or by the use of Minkowski
functionals \cite{meck94} and their combination (the
Shapefinder statistic, see \cite{bhar00}).

\subsection{Global properties: richness, luminosity, mass}
\label{glob}
In order for cluster catalogs to be useful in cosmological studies,
they should also provide cluster mass estimates. Cluster masses can be
estimated directly, by applying dynamical methods (virial theorem, Jeans
equation, \ldots) to the sample of cluster galaxies, if
redshifts are available; by solving the hydrostatic equation for the
X-ray emitting intra-cluster plasma; or by analyzing the effects
of gravitational lensing on background galaxies in the cluster field. 
All these methods are very expensive in terms of
telescope time, hence it is customary to use cheaper global cluster
quantities that can serve as mass proxies. In optical cluster studies,
the most used mass ($M$ hereafter) proxies are the cluster richness
(or multiplicity), $N$, namely the number of galaxies contained in a
cluster, and the cluster luminosity, $L$, both measured in a certain
magnitude range and out to a certain radius from the cluster center.

$N$ and $L$ are evaluated by counting galaxies and, respectively,
summing their luminosities, in a given region of space where we know
the sample is uniformly complete down to a certain magnitude,
$m_c$. If the sample suffers from incompleteness, a correction must be
applied. Both $N$ and $L$ must then be corrected by subtracting the
expected contamination by field galaxies (which can be estimated in a
comparison empty field or from the number counts of general field
galaxy surveys, see, e.g., \cite{lobo97}).  In comparing the $N$ and
$L$ estimates of clusters at different distances, the limiting
magnitudes used must be accordingly scaled.

If one is looking for a mass proxy, the summed luminosity of the
brightest galaxies could suffice. On the other hand, if what is needed
is the {\em total} cluster luminosity, an extrapolation is required to
account for the contribution of the faint galaxies with $m>m_c$. Such an
extrapolation is done by fitting a suitable function to the observed
magnitude distribution, and then integrating this function from $m_c$
to the magnitude of the faintest galaxies.  By far, the most widely
used function is the Schechter LF \cite{sche76}
\begin{equation}
\phi(l)dl=\phi_{\star} \, (l/l_{\star})^{\alpha} \, \exp(-l/l_{\star}) \, d(l/l_{\star}),
\end{equation}
where $l$ is the galaxy luminosity, $l_{\star}$ and $\phi_{\star}$ are
the characteristic luminosity and number density, respectively, and
$\alpha$ is the faint-end power-law exponent. Galaxy luminosities are
derived by converting galaxy apparent magnitudes into absolute
magnitudes via knowledge of the cluster luminosity distance and of the
Galactic extinction in the observational photometric band (see,
e.g., \cite{lin03}). In order to get an estimate of the total cluster
luminosity within a given radius, e.g. the overdensity radius
$r_{200}$, one must determine the luminosity density profile,
Abel-invert it to obtain the 3-d profile (see eq.~(\ref{eq-abel}) in
Sect.~\ref{dynana}), and finally extrapolate it to the desired radius
with a suitable fitting function (e.g. a projected 'NFW' profile
\cite{nava96,mamo05}; see also Sect.~\ref{dyn}).

By the same technique it is in principle possible to get an estimate
of $N(<r_{200})$. However, extrapolation to fainter magnitudes is in
this case dangerous, since faint galaxies largely outnumber bright
galaxies, while their integrated contribution to the total luminosity
is only marginal. A more robust estimate of a cluster richness is
provided by the $B_{gc}$ parameter \cite{yee03}, the galaxy cluster
center correlation amplitude. It is measured by counting galaxies in a
fixed aperture around the cluster center, it requires an assumption
about the shape of the correlation function and the LF, and it must be
corrected for the field galaxy contamination.  $B_{gc}$ is almost
independent of the fixed aperture and chosen limiting magnitude 
\cite{muzz07}.

The most classical direct method to determine $M$, the cluster mass,
is by applying the virial theorem to the projected phase-space
distribution of cluster galaxies (e.g. \cite{binn87}). This method has
been in use since the '30s, when Zwicky and
Smith \cite{zwic33,zwic37,smit36} provided the first preliminary mass
estimates of the Coma and Virgo clusters.  Their studies marked the
discovery of dark matter (DM hereafter; see, e.g. \cite{bivi00}). The
virial theorem is obtained by integrating the equation of hydrostatic
equilibrium for the galaxy distribution in the potential well of a
cluster (the Jeans equation, see \cite{binn87}). In terms of the
observables the virial theorem can be expressed as follows,
\begin{equation}
M=3 \pi \, P \, \sigma_v^2 \, R_h/G,
\label{eq-vir}
\end{equation}
where $G$ is the gravitational constant, $\sigma_v$ is the
line-of-sight velocity dispersion of cluster galaxies, and $R_h$ is
the harmonic mean radius of the projected spatial distribution of
cluster galaxies\footnote{Note that it is customary to use another
  quantity, usually called the ``virial radius'' (see e.g.
  \cite{gira98}), that equals twice the harmonic mean radius. However, the
  radius at a given overdensity, $r_{200}$ or $r_{100}$, is also
  referred to as the ``virial radius''. In order to avoid confusion I
  prefer to use here the harmonic mean radius.},
\begin{equation}
R_h = \frac{1}{2}\frac{n(n-1)}{\sum_{i>j} \, R_{ij}^{-1}},
\label{eq-harm}
\end{equation}
where $R_{ij}$ is the projected distance between two cluster galaxies,
and $n$ is the number of cluster galaxies.  The factor $3 \pi$
corrects for projection effects \cite{limb60,gira98}.  The factor $P$
is the surface pressure term \cite{the86,gira98}.  It is a correcting
factor needed when the entire cluster is not included in the observed
sample. It can be understood as follows. Suppose you have a galaxy
orbiting a cluster with its apocenter at radius $a$ and you observe
the cluster only out to the radius $r_l$, with $r_l<a$. Making the
wrong assumption that the {\em entire} cluster has been observed
corresponds to imposing a smaller apocenter to the galaxy, since $a$
cannot be larger than $r_l$ under this assumption. Given the galaxy
velocity, the same orbital anisotropy, and the same cluster density
profile, imposing a smaller value for the galaxy orbital apocenter
corresponds to forcing a larger mass inside $r_l$.  The correction
factor can be evaluated as follows:
\begin{equation}
P=1-4 \pi r_l^3 \frac{\rho(r_l)}{\int_0^{r_l} 4 \pi x^2 \rho dx}
\, \frac{\sigma_r^2(r_l)}{\sigma^2(<r_l)},
\end{equation}
where $r_l$ is the limiting observational radius, $\rho(r)$ is the
tracer mass density distribution, $\sigma_r$ is the radial component
of the velocity dispersion, and $\sigma(<r_l)$ is the integrated
velocity dispersion within $r_l$ \cite{gira98}. Knowledge of $\rho(r)$
and of the galaxy velocity anisotropy profile is formally needed in
order to solve for $P$; in practice one uses theoretical prejudice in
order to make reasonable assumptions for these profiles. The resulting
$P$ estimate is therefore only an approximation, but this is
preferable to no correction at all, since the correction factor is
systematic, $P<1$ (typically, $P \simeq 0.8$--0.9 for a typical
cluster observed out to $r_l \simeq 2$ Mpc \cite{bivi06}).

Since one is using the velocity and spatial distribution of cluster
galaxies in eqs.~(\ref{eq-vir},\ref{eq-harm}), the resulting $M$
estimate will only be correct if the galaxies are distributed like the
mass \cite{the86,merr87}. Given that different galaxy populations are
distributed differently in projected phase-space (see,
e.g., \cite{bivi02}), the choice of the tracer may change the $M$
estimate rather substantially. Analysis of the mass {\em profiles} of
galaxy clusters have shown that the distribution of ETGs and
red-sequence galaxies is similar to that of the cluster mass (see
Sect.~\ref{mprof}), hence they are to be preferred over late-type,
star-forming (blue) galaxies as tracers of the potential when using
the virial theorem to determine the cluster mass.  Virial masses
estimated using late-type (emission-line) galaxies can be 50\% higher
than those derived from ETGs \cite{bivi97}.

How reliable are the cluster masses derived by applying the virial
theorem to cluster galaxies?  An analysis of clusters extracted from
cosmological numerical simulations has shown that their masses are
only 10\% overestimated by application of the virial theorem to
samples of $n \geq 60$ particles (galaxies) \cite{bivi06}.  Part of
the overestimation is caused by the presence of interlopers, part by
the presence of subclustering. Selecting only ETGs or only relaxed
clusters reduces the bias in the mass estimate. Since much of the
problem lies in the estimate of $R_h$, an alternative mass estimate
based on $\sigma_v$ only can be a viable alternative to the virial
mass estimate \cite{gill04,bivi06}.

Recent analyses of medium-distant clusters (CNOC, EDisCS) have shown
that virial mass estimates based on cluster galaxies are entirely
consistent with mass estimates based on gravitational lensing and/or
on intra-cluster gas X-ray emission
\cite{hick06,john06}, thus confirming the consistency found on local
cluster samples \cite{gira98,wu98}.  Other analyses have found that
virial mass estimates are generally higher than the mass estimates
based on X-ray emission \cite{smai97,wu97,fino01,etto02,ande04,pope05}.  
In general, the largest discrepancies between different mass estimates 
are found for dynamically unrelaxed clusters
\cite{etto02,pope05,cypr06,hash07}.

An indirect confirmation of the reliability of virial mass estimates
has come from the comparison of the values of cosmological parameters
obtained using these mass estimates, with the values obtained by the
Wilkinson Microwave Anisotropy Probe (WMAP \cite{ande04,rine08}).

\subsection{Scaling relations}
\label{scarel}
For a $M$-proxy to be effective, it is important to know precisely its
scaling relation with the cluster mass, as well as the intrinsic
scatter in this relation (see, e.g. \cite{borg01}). The smaller this
scatter, the better is the mass--proxy relation constrained. Moreover,
it is important to know how scaling relations evolve with redshift.
In fact, scaling relations are generally determined locally, on the
best observational samples, yet they have to be applied over a wide
redshift range in order to improve the constraints on cosmological
parameters.

On the cluster scale, both $N$ and $L$ have been shown to be $M$
estimators of similar or even better accuracy than $L_X$
\cite{yee03,pope07,lope06,muzz07}. The typical accuracy with which a
cluster $M$ (within a given overdensity) can be predicted by $N$ or
$L$ is $\sim 30-40$\%
\cite{gira02,bahc03b,yee03,lin04,rame04,mill05,pope07,muzz07}.  At
least half of the observed scatter in the $M-N$ and $M-L$ relations
appears to be intrinsic \cite{muzz07}.  The scatter appears to
increase at lower masses, although it is unclear how much of this
increase is due to observational uncertainties and how much it is due
to an intrinsic larger variance of the properties of low-mass galaxy
systems
\cite{beck07}.  The scatter decreases when irregular and substructured
clusters are removed from the sample \cite{muri02,mill05}.

\begin{figure}
\center{\includegraphics[height=9cm]{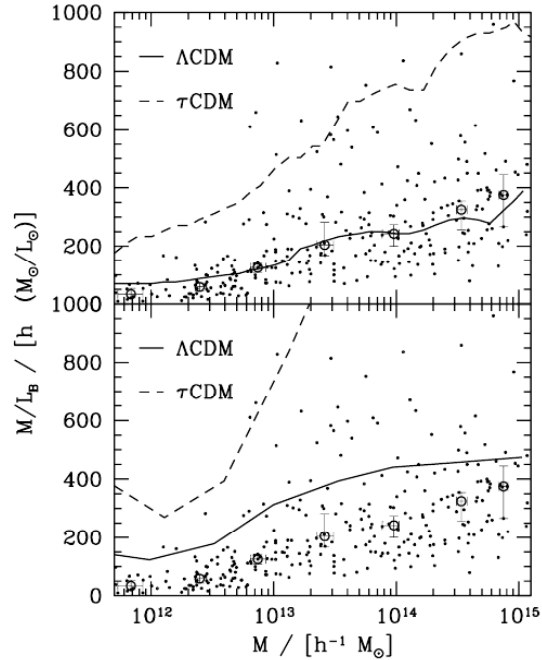}}
\caption{Observed $M/L$ ratio for clusters and groups (dots) and their
median values (circles) with 90\% confidence levels error bars, are
compared with theoretical predictions from two semi-analytical models
(upper panel: \cite{kauf99}, lower panel: \cite{bens00}).
From \cite{gira02}.}
\label{fig-ml}
\end{figure}

The $M-N$ and $M-L$ relations are almost linear on the cluster scale,
but not quite so, and most studies (with some notable
exceptions \cite{bahc03,koch03}) indicate a mild
\cite{scha93,adam98,gira02,lin04,rame04,rine04,pope07,plio06,muzz07} 
or very mild \cite{beck07,koes07b,pope07b} increase 
of the $M/L$ and $M/N$ ratio with cluster
mass, $M/L \propto M^a$ with $a \simeq 0.2 \pm 0.1$.
The $M/L$ ratio is not however well described by a simple power-law
from the cluster to the group mass scale. It steepens considerably for
masses below $\simeq 3 \times 10^{13} \, M_{\odot}$ reaching a minimum
at $\sim 10^{12} \, M_{\odot}$
\cite{gira02,mari02b,eke04b,brou06} (see Fig.~\ref{fig-ml}). 
The non-linearity of the
$M-L$ relation at low masses is however of little concern for most
cosmological studies based on cluster number density, since the
completeness required by these studies is generally achievable only at
the high-$M$ end, at least at high-$z$.

The $M/L$ variation with $M$ in galaxy systems is well fit by the
theoretical predictions of semi-analytical models (e.g.
\cite{kauf99}; see Fig.~\ref{fig-ml}) in which the efficiency of
galaxy formation is inhibited by the reheating of cool gas on
small-mass scales, and by the long cooling times of hot gas on
large-mass scales (see, e.g.  \cite{mari02b}.)  Other models that
could produce the observed $M/L$ variation require galaxy merging
and/or destruction to occur with different efficiency in galaxy
systems of different mass. These mechanisms seem however to be ruled
out by the similar shape of the LFs of clusters with different masses
\cite{pope07b}, and by the lack of any significant evolution of the
cluster $N-M$ relation up to $z \sim 1$
\cite{lin06,gilb07,glad07,gilb08}.

In order to understand the possible biases involved in converting a
given proxy into an $M$ estimate, it is useful to compare scaling
relations involving different proxies, or, equivalently, to determine
the relations between them. In particular, it is interesting to
compare optical and X-ray properties of galaxy clusters.

\begin{figure}
\center{\includegraphics[height=7cm]{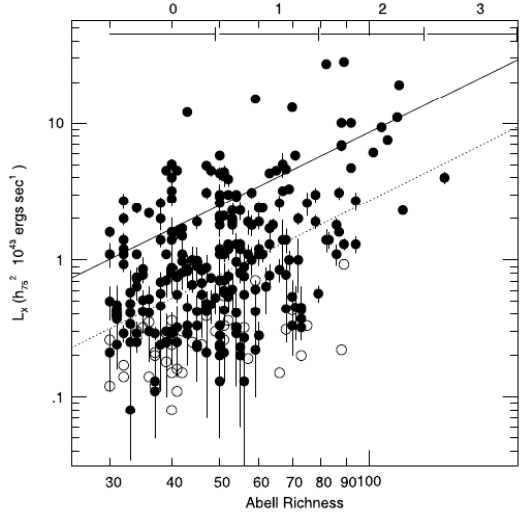}}
\caption{The relation between X-ray luminosity and richness for a
  sample of nearby optically-identified clusters. Filled circles are
  detections, open symbols are upper limits in X-ray luminosity (i.e.
  optically identified clusters undetected in X-ray). The solid
  (dotted) line is the relation that fits the data excluding
  (including) the non-detections.  From \cite{ledl03}.}
\label{fig-nlx}
\end{figure}

In the comparison of X-ray and optical properties of galaxy clusters,
two anomalies emerge: (i) there is a population of optically selected
clusters with too low an X-ray luminosity, $L_X$, for their $\sigma_v$
or optical richness (see Fig.~\ref{fig-nlx};
\cite{bowe97,mcna01,gilb04,lubi04,bark06,fang07,pope07}), and (ii)
there is a population of X-ray bright galaxy groups with too small a
$\sigma_v$ for their X-ray temperature or luminosity
\cite{hels00,osmo04,hels05}.

The X-ray underluminous clusters appear anomalous only with respect to
the $L_X-N$ relation established for {\em X-ray-selected} cluster
samples.  As a matter of fact, low-mass X-ray selected clusters do
follow the same scaling relations established for clusters of higher
masses \cite{valt04}. But X-ray selection excludes low-$L_X$, high-$N$
clusters from the sample, leading to an $L_X-N$ relation that is
biased high relative to the true underlying relation
\cite{dona02,dai07,ryko08}. The true $L_X-N$ relation is characterized
by a large dispersion of the $L_X$ values at given $N$. This can be
partly explained by systematic errors in both the X-ray and optical
estimates (see, e.g., \cite{koce08}). However, part of the dispersion
may be intrinsic and related to different properties of the X-ray
underluminous clusters relative to the normal X-ray emitting cluster
population. X-ray underluminous clusters appear in fact more irregular
\cite{ledl03}, and the properties of their galaxies are reminiscent of
those of infalling field galaxies \cite{pope07}. In other words, the
X-ray underluminous clusters look like unvirialized systems.

While X-ray underluminous clusters may be systems at an {\em early}
stage of their dynamical evolution, X-ray bright groups with
abnormally small $\sigma_v$ may be systems at an {\em advanced} stage
of their dynamical evolution. A possible way to reduce the group
$\sigma_v$ is to slow down galaxies by the process of dynamical
friction \cite{chan43}.  Although the characteristic time of this
process is generally longer than a Hubble time for cluster galaxies,
it is shorter than this for massive group galaxies \cite{hels05}.
Another possibility is that tidal interactions transfer part of the
orbital kinetic energy of group galaxies to their internal energy
\cite{hels05}. It cannot however be excluded that the intrinsic group
$\sigma_v$ are underestimated because of projection, if these groups
have a flattened distribution of galaxies with an anisotropic velocity
dispersion tensor \cite{hels05}.

In order to remove possible biases in the selection of galaxy
clusters, and therefore in the determination of cosmological
parameters, we need to better understand the nature of these outliers
from the relations between optical and X-ray properties.

\subsection{Constraints on cosmological parameters and future surveys}
\label{cosmo}
A traditional way of constraining $\Omega_m$ is by determining the mean
luminosity density, $\rho_L$, and the mean mass-to-light ratio of the
universe, $<M/L>$, via $\Omega_m = <M/L> \times \rho_L/\rho_c$, where
$\rho_c$ is the universe critical density.  This is
known as the Oort technique. This technique works if the systems used to
measure $<M/L>$ are representative of the universe as a
whole. $M/L$ is an increasing function of $M$ at the galaxy and group
scales and flattens at the cluster scales (see Sect.~\ref{scarel}). Hence, the
average $<M/L>$ of rich, massive clusters should be representative of
the universal value. This is also expected from the fact that clusters
are assembled from regions $>10$ Mpc across, so they should contain a
sufficiently large collapsed volume to provide a representative sample of the
average $M/L$ of the universe \cite{muzz07}. Note however that if the
galaxy correlation function on these scales is not unbiased with
respect to total matter, the value for $\Omega_m$ one obtains from the
average $M/L$ of clusters via the Oort method does also depend on
$\sigma_8$, the amplitude of mass fluctuations on 8 $h^{-1}$ Mpc scales
\cite{tink05}.

\begin{figure}
\center{\includegraphics[height=10cm]{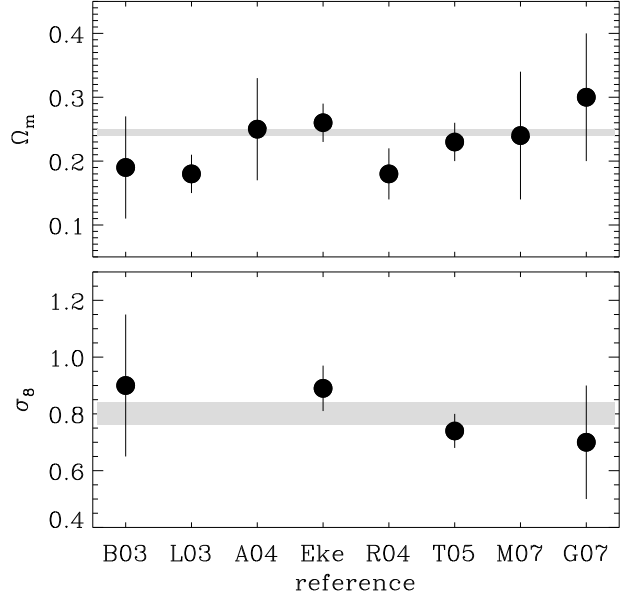}}
\caption{Recent estimates of $\Omega_m$ and $\sigma_8$ obtained via
  the Oort method or via the determination of $n(M,z)$ for
  optically-selected cluster samples. The shaded regions represent the
  WMAP 5-yrs allowed range for the same parameters \cite{dunk08}.
  References for the plotted points are: B03=\cite{bahc03b},
  L03=\cite{lin03}, A04=\cite{ande04}, Eke=\cite{eke04b} (upper panel)
  or \cite{eke06} (lower panel), R04=\cite{rine04}, T05=\cite{tink05},
  M07=\cite{muzz07}, G07=\cite{glad07}.  1-$\sigma$ error bars (as
  given by the authors) are displayed. Note that, when needed,
  $\Omega_m$ has been estimated given $\sigma_8$ from \cite{dunk08} as
  a prior, and vice versa.}
\label{fig-cosmo}
\end{figure}

The first application of Oort's technique probably dates back to 1965
when Abell \cite{abel65} was able to constrain $\Omega_m$ in the
interval $0.1-1.0$ (using number- rather than luminosity-density).
The constraints based on this technique tightened significantly in the
1990's, yielding $\Omega_m \simeq 0.2 \pm 0.1$ and therefore
indicating a low-density universe with high statistical significance
\cite{bahc95,carl96,carl97,adam98b}.

Another way of constraining the cosmological parameters is by
estimating the number density of galaxy clusters (and superclusters),
as a function of their mass, $n(M)$. Constraints on the combination
of the $\Omega_m$ and $\sigma_8$ parameters can be obtained by
comparing the observed $n(M)$ with theoretical predictions (e.g.
\cite{pres74,shet99,jenk01}).  The study of the evolution of $n(M)$
with $z$, $n(M,z)$, can provide even stronger constraints on
cosmological models, allowing to break the $\Omega_m$--$\sigma_8$
degeneracy \cite{borg01b,evra02}.

Recent analyses of optically-selected cluster samples, some based on
the Oort technique, some based on $n(M,z)$, provide estimates for
$\Omega_m$ and $\sigma_8$ in agreement with the values obtained by
WMAP 5-yrs \cite{dunk08} (see Fig.~\ref{fig-cosmo};
\cite{bahc03b,lin03,eke04b,rine04,ande04,tink05,eke06,muzz07,glad07}).

Superclusters can be used to constrain the evolution of cosmic
structure in a more linear regime than applicable to galaxy clusters.
It has been found that the Millennium simulation lacks very rich
superclusters compared to the real universe \cite{eina06}. Similarly,
the existence of a very massive and compact supercluster recently
detected at $z \simeq 0.9$ \cite{gilb08} is a rather unlikely event to
be expected {\em a priori} in the currently favored cosmology.

While these results are encouraging, they are not yet competitive with
those obtained with X-ray cluster surveys (see, e.g., \cite{schu03}).
However, X-ray cluster selection, as well as X-ray cluster mass
estimates may suffer from their own systematics (see, e.g.,
\cite{rasm06,rasi06}). Moreover, at $z>1$ X-ray selection of clusters
does not seem to be as efficient as IR selection \cite{brod08,eise08},
and large samples of optical/IR-selected clusters are expected to come
from ongoing and future very large ($\sim$ thousands of
square-degrees) optical/IR surveys.  Among these, the currently
ongoing Red Sequence Cluster Survey 2,
RCS-2\footnote{\texttt{http://www.rcs2.org/}} is the largest
systematic search for galaxy clusters ever undertaken. It is based on
the deg$^2$ MegaCam imager on the CFHT, it will image $\sim 1000$
deg$^2$ down to $g'=25.3, r'=24.8, z'=22.5$, and will detect clusters
of galaxies up to $z \sim 1$ using the CRS technique. It is estimated
that upon completion, the survey will provide a sample of several
thousand clusters. With such a sample it should be possible to
constrain $\Omega_m$ to an accuracy of $\pm 0.02$ and $\sigma_8$ to an
accuracy of $\pm 0.05$, and to estimate the equation of state of dark
energy, $w$, to within 10\% \cite{yee07}.

Other very ambitious surveys, most of them aimed at the
characterization of the equation of state of dark energy and its
evolution, are currently planned.  They will prove very useful for
distant cluster searches. A four-band survey is planned at the
Panoramic Survey Telescope And Rapid Response System,
Pan-STARSS\footnote{\texttt{http://pan-starrs.ifa.hawaii.edu/public/home.html}},
being developed at the University of Hawaii's Institute of Astronomy.
It will cover 1200 deg$^2$ down to $g=27$ in four bands. The
Kilo-Degree Survey,
KiDS\footnote{\texttt{http://www.astro-wise.org/projects/KIDS/}},
is a 1500 deg$^2$ public imaging survey in the five SDSS bands
that will use the OmegaCAM instrument at the VLT Survey Telescope to
go 2 magnitudes deeper than SDSS. It will be complemented by a near-IR
survey, the VISTA Kilo-degree INfrared Galaxy survey, VIKING.
The Dark Energy Survey,
DES\footnote{\texttt{https://www.darkenergysurvey.org/}} will
cover 5000 deg$^2$ in 5 bands ($g,r,i,z,Y$) with a wide-field
camera to be installed at the Blanco 4~m telescope at CTIO \cite{hons08}.

Not only imaging, also spectroscopic surveys are planned, from which
catalogs of clusters of galaxies will probably be extracted.  The
Baryon Oscillation Spectroscopic Survey, BOSS\footnote{\texttt{http://www.sdss3.org/cosmology.php}}, will observe 10,000
deg$^2$ and obtain redshifts for 1,5 million red luminous
galaxies to $z=0.7$, using the SDSS 2.5m
telescope and spectrographs. BOSS is part of the
SDSS-III and should provide the first data-release in July 2011.
Another planned spectroscopic survey (1 million galaxies in 100
nights) is the Hobby-Eberly Telescope Dark Energy Experiment,
HETDEX\footnote{\texttt{http://hetdex.org}}.

From space, a significant increase in the number of $z>1$ clusters can
come from a mid-IR survey to be conducted with the \texttt{Spitzer}
Space Telescope during its warm mission
\footnote{\texttt{http://ssc.spitzer.caltech.edu}}, expected to last
for about 2 years (see, e.g., \cite{papo08}). In the longer term, an unprecedented amount of data may be provided by two proposed
space-based missions,
ESA's Euclid\footnote{\texttt{http://sci.esa.int/science-e/www/area/index.cfm?fareaid=102}},
the merging of two proposed missions,
DUNE\footnote{\texttt{http://www.dune-mission.net/}} and
SPACE\footnote{\texttt{http://urania.bo.astro.it/cimatti/space/}} (see, e.g., \cite{cima08}),
and
JDEM\footnote{\texttt{http://universe.nasa.gov/program/probes/jdem.html}}
the Joint Dark Energy Mission of NASA and the U.S.  Department of
Energy.
Euclid could provide $\approx 1.5 \times 10^8$ galaxy redshifts
down to $H_{AB}=22$, and a high-resolution, 3-filters, 20,000 deg$^2$
photometric survey of $\geq 2 \times 10^9$ galaxies out to $z \sim
2$.

\section{Structure and dynamics}
\label{dyn}
Early determinations of cluster masses
(e.g. \cite{zwic33,zwic37,smit36}) from application of the virial
theorem to cluster galaxy distributions, implicitly assumed that
galaxies are fair tracers of the cluster gravitational potential, the
so-called {\em light traces mass} hypothesis. However, the result of
these analyses did not provide support for this assumption, since the
derived masses were orders of magnitude larger than the sum of the
masses of the visible galaxies. What if galaxies were not distributed
like the total mass? Virial mass estimates could be biased high or
low \cite{the86,merr87}.  Comparison with other cluster mass
estimates \cite{gira98,wu98}, and analyses of simulated clusters
extracted from cosmological simulations
\cite{bivi06} have since proven that the virial mass estimates are more
or less correct. This then suggests that the projected spatial and
velocity distribution of cluster galaxies is not very different from
that of the total mass. However, {\em proving} this is not so simple,
one must compare the distribution of the tracer population to the
distribution of the total mass (see Sect.~\ref{mlr}).

Knowing the mass distribution within clusters (also in relation to the
distribution of the different cluster components) not only is
important for a correct estimate of cluster masses, but also because
it provides important clues on the formation and evolution of galaxy
clusters and their components (e.g. \cite{spri01,elza04,gao04}), and on
the nature of DM (e.g. \cite{sper00,mene01,reed05}). E.g. warm DM
is expected to produce lower density halo cores than cold DM (CDM). If
DM is cold, halos should be characterized by density profiles with 
a central cusp, such as the NFW profile \cite{nava97},
\begin{equation}
  \rho_{NFW}=\frac{\rho_0}{(c r/r_{200})(1+c r/r_{200})^2},
\end{equation}
where $c$ is the concentration parameter.
Modifications of the NFW profile have been suggested, all
characterized by the central cusp \cite{moor99,haya04,diem05}.
Another widely used cuspy density profile is the Hernquist model
\cite{hern90},
\begin{equation}
\rho_{Hernquist} = \frac{\rho_0}{r \, (r+r_H)^3}. 
\end{equation}
On the other hand, observations of galaxy
rotation curves have revealed the presence of a central core
(e.g. \cite{debl02,borr03,debl03,gent04,debl08}). 
Cores could be created in the mass distribution
if the DM particles are self-interacting \cite{sper00}, or if
galaxies are able to pump energy into the DM component via dynamical
friction \cite{chan43,elza01}. 
Cored mass density models have been suggested
\cite{burk95,arie03}, such as the Burkert profile,
\begin{equation}
  \rho_{Burkert}=\frac{\rho_0}{(1+r/r_c)[1+(r/r_c)^2]},
\end{equation}
characterized by the core radius $r_c$.
Another widely used mass profile is
the softened isothermal sphere,
\begin{equation}
\rho_{SIS} = \frac{\rho_0}{1+(r/r_c)^2}.
\end{equation}
Note that at large radii $\rho_{Burkert} \propto r^{-3}$ like the NFW
profile, while $\rho_{SIS} \propto r^{-2}$.

Given the problems that the NFW model has at small scales, it is
important to test its validity on larger scales, i.e. on cluster- and
group-sized halos. Hence determination of the cluster mass profile,
$M(r)$, becomes a crucial test for the CDM cosmological model (see
Sect.~\ref{mprof}).

\subsection{Dynamical analysis: methods}
\label{dynana}
The most commonly used method to determine the mass profile $M(r)$ of
galaxy clusters is the Jeans method (see \cite{binn87}) hereafter
described.

Assuming spherical symmetry, the projected number density profile
$N(R)$ of a tracer can be uniquely deprojected via the Abel inversion
equation \cite{binn87},
\begin{equation}
\nu(r) = - \frac{1}{\pi} \int_{r}^{\infty} \frac{dN}{dR} 
\frac{dR}{\sqrt{R^2-r^2}},
\label{eq-abel}
\end{equation}
where $\nu(r)$ is the 3-d number density profile, $R$ and $r$ are the
projected and, respectively, the 3-d radius (i.e. the distance from
the cluster center). While this deprojection is straightforward,
deprojecting the line-of-sight velocity dispersion profile,
$\sigma_v(R)$, requires knowledge of the velocity anisotropy profile,
\begin{equation}
\beta(r) \equiv 1 - \frac{\rm{<} v_t^2 \rm{>}(r)}{\rm{<} v_r^2 \rm{>}(r)},
\label{eq-beta}
\end{equation}
where $\rm{<} v_r^2 \rm{>}(r)$, $\rm{<} v_t^2 \rm{>}(r)$ are the mean
squared radial and tangential velocity components, which we can
write as $\sigma_r^2(r)$ and $\sigma_t^2(r)$ respectively, in the absence
of bulk motions and net rotation. In the simplest case of the
isotropic velocity distribution, $\beta(r) \equiv 0$, the $\sigma_v$
deprojection reads
\begin{equation}
\sigma_r^2 (r) = - \frac{1}{\pi \nu(r)} \int_{r}^{\infty}
\frac{d [N \, \sigma_v^2]}{dR} 
\frac{dR}{\sqrt{R^2-r^2}}.
\end{equation}
Through a more complicated set of equations it is possible to
deproject $\sigma_v$ in the case of generic $\beta$ \cite{mamo08}.

Given $\sigma_r(r)$ and $\beta(r)$ it is possible to determine the
mass profile, $M(r)$, through the Jeans equation for a collisionless
system of particles (e.g. galaxies),
\begin{equation}
M(r) = - \frac{r \sigma_r^2}{G} \left( \frac{d
\ln \nu}{d \ln r} + \frac{ d \ln \sigma_r^2}{d \ln r} +2 \beta \right).
\end{equation}
Similarly, given $M(r)$ and $\beta(r)$ it is possible to determine
the observed, projected phase-space distribution of the tracers via
\begin{equation}
\nu(r) \, \sigma_r^2(r) = G \int_{r}^{\infty} 
\frac{\nu M}{\xi^2} \, \exp \left[ 2 \int_{r}^{\xi} \frac{\beta 
dx}{x}\right] d \xi,
\end{equation}
and
\begin{equation}
N(R) \sigma_v^2(R) = 2 \int_{R}^{\infty} \left( 1 - \beta \frac{R^2}{r^2}
\right) \frac{\nu \, \sigma_r^2 \, r \, dr}{\sqrt{r^2-R^2}}.
\end{equation}
From the eqs. above it is clear that the same observed number density
and velocity dispersion profiles $N(R)$ and $\sigma_v(R)$ can be
obtained by a different combination of the mass and anisotropy
profiles $M(r)$ and $\beta(r)$, and vice versa. Symmetrically, given
the observables $N(R)$ and $\sigma_v(R)$ it is possible to obtain
$M(r)$ only if $\beta(r)$ is known and $\beta(r)$ only if $M(r)$ is
known \cite{mamo08,binn82,sola90,dejo92}.  This is the so-called
``mass--anisotropy'' degeneracy.

In order to break this degeneracy, one must constrain $\beta(r)$
independently from $M(r)$. A possibility is to build distribution
function models (see, e.g., \cite{vdma00}) and use them to compute the
probability that a particle observed at a given projected radius $R$
have a line-of-sight velocity $v$ in a given interval $dv$. These
probabilities are then used in a maximum likelihood analysis to
determine the model that best represents the observed projected
phase-space distribution of galaxies. The best-fit model can also be
chosen by comparing the line-of-sight velocity
distribution predicted by the model with the observed histogram of
velocities of cluster galaxies. The comparison in this case can be
done by considering moments of the velocity distribution higher
than the second (e.g. \cite{merr87,vdma00,loka03,wojt05}).
Robust estimates of these moments are obtained by the use of 
Gauss-Hermite polynomials.

Another way to break the mass--anisotropy degeneracy is to consider
several independent tracers of the gravitational potential, then apply
the Jeans procedure independently for each of the
tracers. Subject to the constraint that different tracers should
provide identical $M(r)$ solutions, it is possible to reduce the range
of acceptable $[M(r),\beta(r)]$ (see, e.g., \cite{batt08}).

The Jeans procedure outlined above also assumes that the system is in
dynamical equilibrium.  However, since clusters grow by accretion of
field galaxies (e.g. \cite{moss77,bivi97}) they are not steady-state
systems.  One should then include the time derivative term in the
Jeans equation (eq. 4-29c in \cite{binn87}). Fortunately, the rate of
mass accretion onto low-$z$ clusters is small, $\simeq 8$\% of
the total mass in a dynamical time ($\sim 1/10$ the Hubble time)
\cite{adam05}, although it increases with redshift \cite{elli01}.
Since the accretion process is not smooth, some clusters, even at
low-$z$, may be observed during an intense accretion phase. These
clusters are nevertheless easy to spot, since most of the accreted
mass is in the form of groups \cite{zabl93} which can be identified as
substructures in the projected phase-space distribution of cluster
galaxies (see Sect.~\ref{subs}). More problematic is the case of
galaxy groups \cite{giur88} and of high-$z$ galaxy clusters, since
several of these systems are likely to be detected when they are still
in their collapse phase and far from dynamical equilibrium.

Other usual assumptions of the Jeans analysis are sphericity and the
absence of net rotation. Deviation from spherical symmetry has been
shown not to be a major problem for individual clusters
\cite{vdma00,sanc04} and there is little if any evidence for net
rotation in galaxy clusters \cite{hwan07}.

The Jeans equation applies to a collisionless system of particles.
Galaxies do behave as quasi-collisionless particles when they move at
high speed, which is the case in galaxy clusters. High-speed galaxy
encounters produce little tidal damage and do not lead to mergers
\cite{mamo95}, and only the most massive galaxies have their motions
slowed down by dynamical friction \cite{bivi92}. As the mass of the
host system of galaxies decreases, dissipative processes become more
important.  Groups in particular, are very favorable sites for galaxy
mergers \cite{osmo04} so the collisionless Jeans
equation may not be applicable for these systems \cite{menc96}.

Interlopers are another serious problem when one is trying to
determine a cluster mass profile. Interlopers are
foreground/background galaxies that happen to lie in the same
projected phase-space region occupied by cluster galaxies. Several
methods exist to get rid of them. Tests on cluster-scale halos
extracted from cosmological simulations have shown these methods to
perform relatively well (see in particular
\cite{bivi06,wojt07,wojt07b}).  There are however interlopers that are
impossible to distinguish from real cluster galaxies; in order to deal
with these interlopers a statistical approach is generally adopted
(e.g. \cite{wojt07}).  In the statistical approach the projected
phase-space distribution of galaxies observed in the cluster region is
assumed to be contaminated by a certain fraction of interlopers with a
well known spatial and velocity distribution (inferred from the
analysis of simulated halos).  Alternatively, one can use galaxy
internal properties, such as, e.g., colors, spectral types, and
morphologies, to improve the separation between cluster and field
galaxies.

In order to determine a reliable cluster mass from the projected
phase-space distribution, $\geq 60$ galaxies are needed \cite{bivi06},
but about an order of magnitude more are required for the
determination of a cluster mass {\em profile} (see, e.g.,
\cite{loka06}).  Spectroscopic samples of several hundred member
galaxies per cluster are still rare, hence it is common practice to
build a 'composite' cluster by stacking together the data of several
clusters (see, e.g., \cite{carl97,vdma00,bivi03,katg04,loka06b}). The
projected phase-space distributions of different clusters can be put
together by scaling galaxy radii and velocities with virial quantities
($r_{200}, v_{200}$\footnote{The so-called circular velocity $v_{200}$
is defined as $v_{200}=(G M_{200}/r_{200})^{1/2}$ where $M_{200}=(4 \pi /3) \,
200 \rho_c \, r_{200}^3$.}). This procedure
is supported by the results of cosmological numerical simulations,
that suggest that halos at the cluster mass scale are a
quasi-homologous family of objects, their mass profiles changing only
slightly with the halo mass
\cite{nava97,dola04}. 

By stacking clusters together it is possible to deal with samples of
a few thousand cluster galaxies. These samples are obtained from
cluster-dedicated spectroscopic surveys, such as ENACS
\cite{katg96,katg98} and the CNOC \cite{yee96,elli98}, and also from field
spectroscopic surveys wherein clusters have been identified, such as
the 2dFGRS \cite{coll01,depr02} and the SDSS (see e.g.
\cite{mill05,rine06}).  The most recent cluster-dedicated
spectroscopic surveys are the Las Campanas/Anglo­Australian Telescope
Rich Cluster Survey (LARCS \cite{pimb01}), the Cluster and Infall
Region Nearby Survey (CAIRNS \cite{rine03}), the WIde Field Nearby
Galaxy-clusters Survey (WINGS \cite{fasa06}), and the
EDisCS \cite{whit05}.

In recent years, another technique, usually referred to as the
Caustic method (\cite{diaf97,diaf99}) has been developed. In
this method, a cluster $M(r)$ is obtained from the amplitude of the
caustics delineated by the projected phase-space distribution of
galaxies in the cluster region.  This amplitude is related to the
gravitational potential through a function ${\cal F}$ of the projected
radius, $R$, of the gravitational potential, and of $\beta(r)$ (see eqs. 9 and
10 in \cite{diaf99}). Also this method suffers from the
mass--anisotropy degeneracy, but only in the central cluster
regions, since numerical simulations indicate ${\cal
F} \approx \rm{const}$ at large radii,
$R>r_{200}$ (see Fig.~2 in \cite{diaf99}). Hence the Caustic
mass estimate is relatively robust at large radii, exactly where
the Jeans estimate may be more affected by problems of
interlopers and deviations from dynamical equilibrium.  On the other
hand, the Jeans method is more robust at small radii, where
imperfect knowledge of ${\cal F}$ increases the systematic
uncertainty in the Caustic mass estimate.

The Caustic and Jeans methods have been shown to
produce consistent cluster mass profiles \cite{bivi03,rine03}.  Through
analyses based on samples of cluster-sized halos extracted from
cosmological numerical simulations, both methods have been shown to be
reliable \cite{diaf99,sanc04,loka06}. There are only a few direct
comparisons of mass profiles determined from the distribution of
cluster galaxies with those determined using the X-ray-emitting
intra-cluster gas or via gravitational lensing. In general, lensing
mass profiles are in agreement with those determined via the
Caustic \cite{diaf05} and Jeans \cite{nata96}
methods. On the other hand, the agreement is less good between X-ray
determined mass profiles and those determined from the distribution of
cluster galaxies \cite{diaf05,bena06}.

\subsection{Mass profiles}
\label{mprof}
Application of the Caustic technique to the CAIRNS sample
\cite{gell99,rine00,rine01,rine03,rine04} has shown that the cluster
mass density profile $\rho(r)$ resembles the NFW model, except at
large radii, $r> 2 \, r_{200}$, where its slope seems to be somewhat steeper
(between $-3$ and $-4$). SIS models are rejected. The best-fit values
of the NFW concentration parameter range between 5 and 17. Very
similar results have recently been obtained by application of the same
Caustic technique to a new sample of 72 X-ray selected clusters
extracted from the SDSS (the CIRS sample \cite{rine06}).

\begin{figure}
\center{\includegraphics[height=7cm]{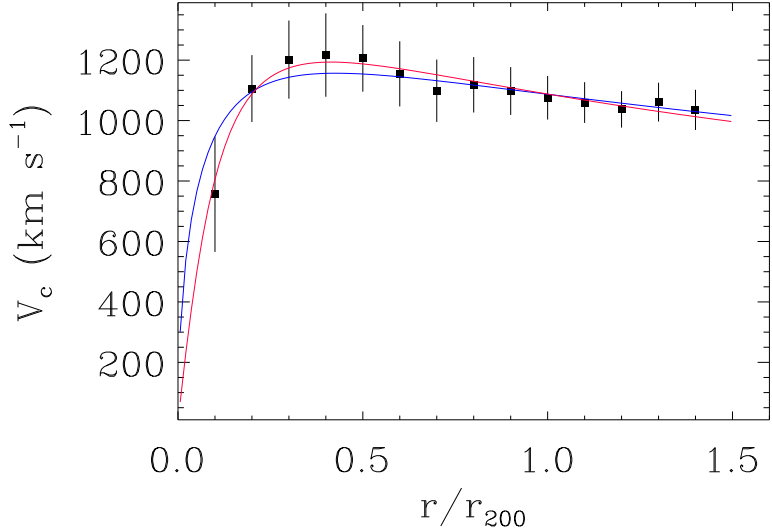}}
\caption{The DM circular velocity profile ($V_c=\sqrt{GM/r}$) of the
  stacked cluster from the ENACS data-set \cite{bivi06b}
  (points with 1-$\sigma$ random $+$ systematic error bars).  The
  best-fitting NFW and Burkert models are indicated with a blue and
  red line, respectively.}
\label{fig-vcirc}
\end{figure}

A combination of the Jeans and Caustic analysis was used to determine
the mass profile of a composite cluster extracted from the 2dFGRS
\cite{bivi03}. By stacking together 43 nearby clusters, a total sample
of 1345 cluster galaxies was obtained. Late-type galaxies (LTGs in the
following) were excluded from the sample, and isotropy was assumed for
the Jeans analysis.  The resulting $M(r)$ was found to be well
described by a $c \simeq 6$ NFW profile over the radial range 0--2
$r_{200}$.  If a cored profile is fitted to $M(r)$, the core radius is
constrained to be small, $< 0.1 \, r_{200},$ i.e.  not much larger
than the size of the BCG which generally sits at the cluster center.

A composite of 1129 ETGs was constructed from 59 clusters from
the ENACS sample \cite{katg04}.  By comparing the velocity
distribution of cluster ETGs with distribution function models,
stringent constraints on $\beta$ were obtained. The ETGs were
shown to move on nearly isotropic orbits, hence $\beta \equiv 0$ was
adopted in the Jeans analysis. It was found that
$\rho \propto r^{-2.4 \pm 0.4}$ at $r=r_{200}$, fully consistent with
the NFW asymptotic slope.
Two models were shown to provide an adequate fit to the
data, a NFW profile with $c=4 \pm 2$, and a Burkert profile with a
rather small core radius ($\leq 0.1 \, r_{200}$).
The $M(r)$ solution obtained by using ETGs as
isotropic tracers was later confirmed by using another tracer of the
gravitational potential, i.e. cluster Sa--Sb galaxies \cite{bivi04}.

The mass distributions of a few individual nearby clusters (including
Coma) have also been determined \cite{loka03,loka06,wojt07b}.  In
order to break the mass-anisotropy degeneracy not only the projected
velocity dispersion profile but also the velocity kurtosis profile
were derived and modeled.  All cluster mass profiles turned out to be
well described by an isotropic NFW model, with a median$(c)=7$.

The ENACS data-set was re-analyzed to determine the relative
contributions of baryons and DM to the total mass profile
\cite{bivi06b}. Since the DM contribution is dominant, the resulting
DM $M(r)$ does not differ significantly from the total mass $M(r)$, it
is only slightly more concentrated (NFW $c=5 \pm 1$, Burkert $r_c=0.12
\pm 0.02$; see Fig.~\ref{fig-vcirc}). If the subhalos DM contribution
is subtracted from the whole DM component, what is left is the diffuse
DM associated with the main halo (cluster), which appears to be even
more concentrated (NFW $c=7 \pm 1$, Burkert $r_c=0.09 \pm 0.02$). Note
however that splitting the DM $M(r)$ into its halo and subhalo components
is very model-dependent.

\begin{figure}
\center{\includegraphics[height=7cm]{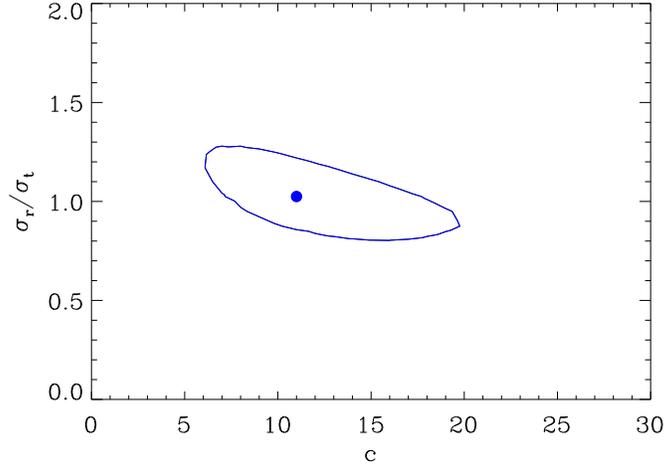}}
\caption{The velocity anisotropy $\sigma_r/\sigma_t$ vs. the
NFW concentration $c \equiv r_{200}/r_s$ parameter as determined for a
stacked sample of 26 X-ray-emitting small groups, mostly from the GEMS
sample \cite{osmo04}.  1-$\sigma$ contours are shown \cite{bivi09}.}
\label{fig-gems}
\end{figure}

These results show that central cuspy models such as NFW and Hernquist
provide an adequate fit to the mass profile of nearby galaxy clusters.
The Burkert profile is also acceptable, as far as the core radius is
small, of order the size of the BCG or smaller. Since galaxies are
treated as point-like tracers of the potential in the Jeans analysis,
the size of the core radius is close to the resolution size of the
analysis. The upper limit on the size of the core radius can be used
to constrain the DM scattering cross section by comparison with
simulations \cite{mene01}. The resulting upper limit, $<2$ cm$^2$
g$^{-1}$, effectively rules out Self-interacting DM as a possible way
to explain the cored mass density profile of dwarf galaxies
\cite{sper00,dave01}. The absence of a significant core in the cluster
mass distribution also implies that dynamical friction is not very
effective in transferring cluster galaxy kinetic energy to DM
particles \cite{elza01}.  This is consistent with observational
estimates of galaxy luminosity segregation in clusters 
(see, e.g., \cite{bivi02}).

While the inner slope of the density profile is essentially
unconstrained, at large radii the asymptotic slope of the density
profile is constrained to lie between $-3$ and $-4$, consistent with
the NFW, Hernquist, and Burkert models, but not with the SIS model.

At higher redshifts the constraints are less strong. Results based on
the analysis of 16 stacked CNOC clusters confirm that NFW is an
acceptable mass profile model on cluster scales also at $0.17 \leq z
\leq 0.55$ \cite{carl97,carl97b,vdma00}.

\begin{figure}
\center{\includegraphics[height=7cm]{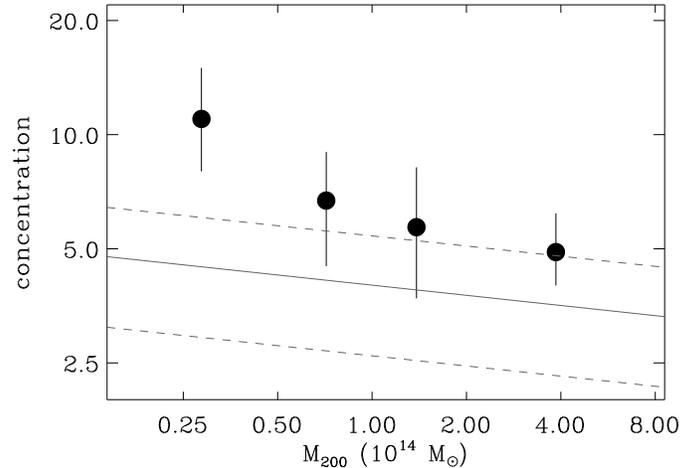}}
\caption{The NFW concentration $c \equiv r_{200}/r_s$ parameter vs.
  the virial mass $M(r_{200})$ as determined on four samples of groups and
  clusters \cite{bivi09,mahd99,bivi03,bivi06b} characterized by different
  average masses (points with 1-$\sigma$ error bars). The solid and
  dashed lines indicate the median and $\pm 1$-$\sigma$ predictions for
  a $\Lambda$CDM cosmology with WMAP5 cosmological parameters
  \cite{duff08}.}
\label{fig-cm}
\end{figure}

The CDM-motivated NFW model fits well the mass profiles of
cluster-sized halos, and does not fit the mass profiles of
galaxy-sized halos. Hence it is important to test the model at
intermediate scales (groups of galaxies). Unfortunately, the results
for the group $M(r)$ are still controversial so far. From the
analysis of 588 galaxies in 20 stacked groups the group density
profile was found to be consistent with the Hernquist
model \cite{mahd99}, but using a sample twice as large the same authors
concluded that a single power-law model is a better representation of
the data \cite{mahd04}.  Both results are inconsistent with those obtained
for a higher redshift group sample ($0.1 \leq z \leq 0.55$) whose
average mass density profile is characterized by a very shallow-slope
and a central core \cite{carl01}.

Discrepant results probably arise as a consequence of different
selections of the group samples, since not all groups are dynamically
virialized systems \cite{giur88,diaf93,mamo95,mahd99}.  A proper
characterization of the group mass profile awaits a careful definition
of the group sample, based on the group characteristics.  Such a
sample may be provided by the groups of the Group Evolution
Multiwavelength Study (GEMS \cite{osmo04}) for which both X-ray and
optical data are available.  Comparing optical and X-ray group
properties helps constraining the group dynamical status
\cite{brou06}. A preliminary analysis of this sample suggests
consistency of the average group mass profile with the NFW model, with
a higher concentration parameter than for clusters, in line with
predictions from $\Lambda$CDM (see
Fig.~\ref{fig-gems}; \cite{bivi08,bivi09}).

In Fig.~\ref{fig-cm} results for the best-fit NFW concentration parameter are
collected from the literature. The $c=c(M)$ trend is somewhat above
recent predictions from $\Lambda$CDM cosmological numerical
simulations \cite{duff08}.

\subsection{The relative distribution of dark and baryonic matter}
\label{mlr}
In order to answer the question raised in Sect.~\ref{dyn}, i.e. are
galaxies distributed like the DM, the mass density profiles of galaxy
systems must be compared with their galaxy {\em number-} or {\em
luminosity-}density profiles.  These can be evaluated by counting galaxies
or, respectively, summing galaxy luminosities, in concentric annuli
around the cluster center, taking into account the completeness
correction, if needed. The mass-density to luminosity- (or number-) density
profiles ratio is called the mass-to-light ($M/L$) profile.

Since different cluster galaxy populations have different
distributions (see Sect.~\ref{morph}) there is not a unique $M/L$ profile.
Depending on the photometric band used to select the cluster galaxies,
the relative fraction of red and blue, quiescent and star-forming,
early- and late-type galaxies in the resulting
sample may vary. Modulo the selection in type or color, the $M/L$
profiles found by different authors are generally in agreement
\cite{bivi03,katg04,rine04,bivi06b}. The cluster $M/L$ profile increases
from the center to $0.2 \, r_{200}$, flattens out to $r_{200}$, and
then decreases, by a factor $\times 2$ out to the turnaround radius.
The trend near the center is caused by the presence of the BCG which
sits at the bottom of the cluster potential well. The external,
decreasing trend is instead caused by the increasing fraction of
late-type, blue galaxies with radius.  Selecting only the red,
early-type, quiescent cluster galaxies (or selecting galaxies in the
$K$ band) flattens the $M/L$ profile in the outer parts. Removing the
BCG flattens the $M/L$ profile near the center. Hence, the
light of red cluster galaxies (except the brightest one) does indeed
trace the mass, but the light of blue cluster galaxies does not. 
Applying the virial theorem to the distribution of red cluster galaxies
(BCG excluded) should then provide unbiased mass
estimates for dynamically relaxed clusters \cite{bivi06}.

\begin{figure}
\center{\includegraphics[height=9cm,width=10cm]{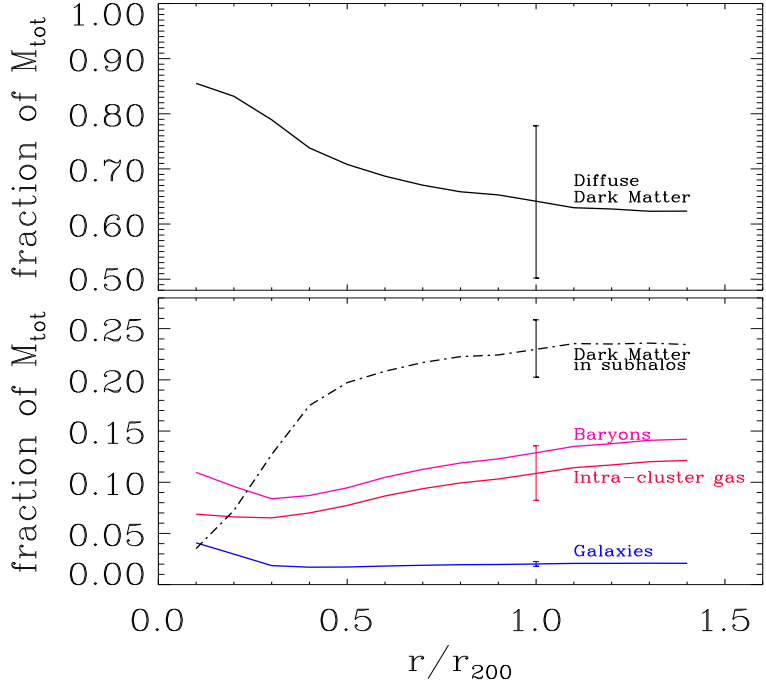}}
\caption{The fractional mass distributions of the different components
  of galaxy clusters, i.e. diffuse DM, DM associated with subhaloes,
  and baryons (intra-cluster gas plus galaxies)
  \cite{bivi06b}. Representative 1-$\sigma$ (random $+$ systematic)
  error bars are indicated.}
\label{fig-mfrac}
\end{figure}

While galaxies are useful tracers of the cluster potential, they are
by far a negligible component not only of the cluster mass, but also
of the baryonic mass. Most of the baryonic mass is in the
intra-cluster, X-ray emitting gas. The intra-cluster gas-to-total mass
fraction increases with radius as $r^{0.4}$ \cite{bivi06b}. Hence, the
baryonic mass is less concentrated than the total mass at all radii,
except near the very center, where the baryons of the BCG dominate the
mass budget (see Fig.~\ref{fig-mfrac}).

Exploring the evolution of the $M/L$ profile with redshift can provide
useful information on when and how the different galaxy populations
settle in galaxy clusters. The poor constraints existing so far for
clusters at $z \approx 0.3$ seem to confirm that $M/L \approx$
constant out to the virial radius, when only red galaxies are selected
\cite{vdma00}.

Since the average mass profile of low-mass galaxy systems (groups) 
is not yet well constrained, results on their $M/L$ profile are
controversial \cite{mahd99,carl01}.
Cosmological numerical simulations indicate that
groups have {\em more} concentrated mass profiles than clusters (see e.g.
\cite{nava97,dola04}), while observations indicate that groups have {\em
  less} concentrated galaxy number density profiles than clusters
\cite{pope07b}, hence groups might be characterized by steeper $M/L$ 
profiles than clusters.

\subsection{The orbits of galaxies and mass accretion}
\label{accrete}
According to the hierarchical model for the formation and evolution of
cosmic structures, clusters grow from accretion of galaxies and galaxy
groups from the field. CDM cosmological numerical simulations have
shown that DM particles accrete onto clusters on moderately radially
elongated orbits, i.e. with a radial velocity anisotropy that
increases moving out to the virial radius (e.g.
\cite{torm97,ghig98,diaf99,diem04}).  Observational evidence
supporting the hierarchical build-up of galaxy clusters has been
provided by the discovery that cluster ETGs and LTGs have different
kinematics \cite{moss77,sodr89,bivi92,carl97,bivi97}. LTGs are
characterized by a larger $\sigma_v$ than ETGs, and this has
been interpreted as evidence that LTGs are an infalling, unvirialized
population.  However, kinematical evidence alone cannot prove the LTGs
are indeed an infalling population, full dynamical modeling is
required (see Sect.~\ref{dynana}).

\begin{figure}
\center{\includegraphics[height=7cm,width=9cm]{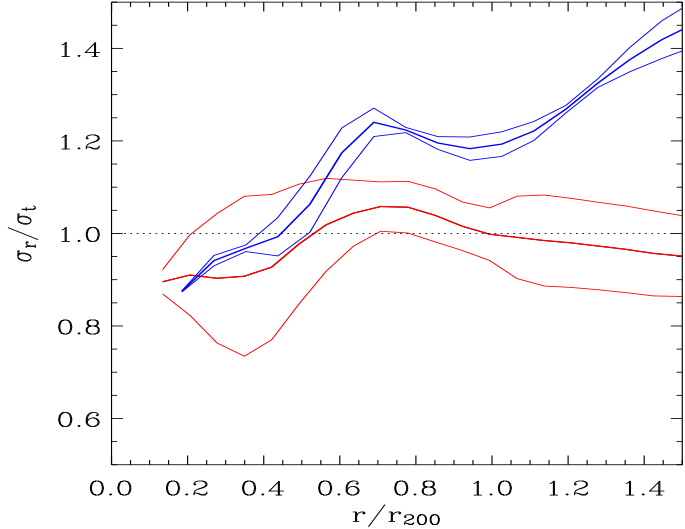}}
\caption{A preliminary determination of the velocity anisotropy
  profiles of the red and blue galaxies in the CIRS clusters
  and their $\pm 1$-$\sigma$ confidence levels \cite{bivi09b}.}
\label{fig-anis}
\end{figure}

One of the first full dynamical modeling of a cluster was made in the
early 80s for the Coma cluster \cite{kent82}. It was concluded that
the galaxy orbits are not primarily radial. Consistently, many recent
dynamical modeling of low-$z$ galaxy clusters, mostly based on stacked
cluster samples from the ENACS, CNOC, and SDSS data-sets, have
concluded for quasi-isotropic orbits of ETGs
\cite{carl97b,mahd99,bivi02b,loka03,katg04}.  Since ETGs are the
dominant cluster galaxy population, also the mean velocity anisotropy
(see eq.~\ref{eq-beta}) of cluster galaxies altogether is found to be
$\beta \approx 0$ \cite{vdma00,rine03,loka06}. Interestingly, full
dynamical analysis does not support the interpretation of LTGs as an
unvirialized infalling population.  Probably the details of the
results depend on how accurately are interlopers rejected from the
sample of cluster members. Anyway, both for nearby and medium-$z$
clusters LTGs are found to be in dynamical equilibrium within the
cluster potential.  At variance with ETGs, however, LTGs have
moderately radially anisotropic orbits
\cite{carl97b,mahd99,bivi02b,bivi04,hwan08}, with an anisotropy that
increases with radius \cite{bivi04,bivi08,bivi09b} (see
Fig.~\ref{fig-anis}).  A finer distinction of the LTG population into
two classes, Sa--Sb on one side and Sbc--Irr on the other, has shown
that the radial anisotropy is characteristic of the latter class
only, while the orbits of Sa--Sb are isotropic within
observational uncertainties \cite{bivi04}.

The velocity anisotropy profile of LTGs is remarkably similar to that
of DM particles in clusters extracted from cosmological numerical
simulations (see, e.g., \cite{ghig98,diaf99,diem04}). The orbital
characteristics of DM particles are reminiscent of their almost radial
accretion onto clusters along the surrounding filaments. By analogy,
also the predominantly radial orbits of LTGs can be taken as an
indication that these galaxies retain the dynamical memory of their
infalling motions into the clusters. They are probably newcomers of
the cluster environment, where they have spent too little time for the
dynamical memory of their initial infall to be totally erased.  ETGs,
on the other hand, have had time to undergo sufficient energy and
angular momentum mixing, capable of isotropizing their orbits (see,
e.g., \cite{lu06} and references therein).  Such energy and angular
momentum mixing occurs in galaxy systems mostly via phase- and 
chaotic-mixing or violent relaxation
\cite{heno64,lynd67,kand03,merr05,lapi08}, which occur when the system
gravitational potential changes rapidly, i.e. at the time of the
system assembly or on the occurrence of major mergers
\cite{manr03,peir06,vall07}. Another process capable of isotropizing
galaxy orbits is the secular growth of cluster mass \cite{gill04}.

LTGs have probably entered the cluster environment after the last
major merger. Consistent with this hypothesis is the fact that they
still retain most of their gas content, which cluster-related
environmental processes will eventually strip given sufficient time
(see Sect.~\ref{processes}). Numerical simulations confirm that
recently accreted satellites in a host halo have more radially
extended and less bound orbits \cite{tayl04,tayl05}.

Independent, direct evidence for the accretion of field spirals (S)
onto clusters has been obtained from the analyses of the
distance--velocity diagram around the Virgo clusters and other nearby
galaxy systems \cite{tull84,gava91,cecc05}. Unfortunately, distance
measurements are affected by large uncertainties and cannot yet be
used to assess the infall process in a statistically significant
sample of massive clusters.

There is no direct estimate of the evolution with $z$ of the ETG and,
separately, the LTG $\beta(r)$. However, an indirect argument can be
used to rule out significant evolution at least up to $z \simeq 0.3$.
The projected phase-space distributions of ETGs and, separately, LTGs
in the $z \simeq 0.1$ ENACS clusters and in the $z \simeq 0.3$ CNOC
clusters are remarkably similar \cite{bivi08}, except for the
normalization (the fraction of blue galaxies is higher in more distant
clusters, the so-called Butcher-Oemler effect -- see
Sect.~\ref{colors}). Also the average $M(r)$ of ENACS and CNOC
clusters are very similar \cite{katg04,vdma00}.  Similarity of the
observed projected phase-space distributions and of the mass profiles
then imply, from the Jeans analysis (see Sect.~\ref{dynana}),
similarity of the velocity anisotropy profiles. This implies, in
particular, that, just like their low-$z$ counterparts, also the CNOC
LTGs are newcomers of the cluster environment.  Given that their
fraction decreases with time, cluster LTGs must either
transform in ETGs or dim with time (or both). If they do undergo color
and morphological transformation, they need at the same time undergo
orbital isotropization since ETG orbits are isotropic and LTG orbits
are moderately radial. This observation could prove useful in
constraining the mechanisms that drive galaxy evolution in clusters
(see Sect.~\ref{processes}).

The mass accretion rate can be estimated by measuring the mass outside
the virial cluster region which is bound to infall into the cluster in
the future. Using the Caustic method it has been estimated that
$M(r<r_{200}) \approx M(r_{200} <r<r_t)$, where $r_t \approx 5 \,
r_{200}$ is the turnaround radius, i.e. the characteristic radius that
separates accretion from outflow regions \cite{rine06}.  According to
$\Lambda$CDM cosmological N-body simulations, clusters will reach
their final $M_{200}$ mass when they are 32 Gyr old \cite{bush05}.
The Caustic estimate then implies a future average accretion rate of
$\dot{M} \approx 0.06 \, M_{200}/\mbox{Gyr}$. This estimate is in
excellent agreement with an independent estimate obtained by summing
up the mass of recently accreted groups in the Coma cluster
\cite{adam05}, $\dot{M} \simeq 0.02-0.11 \, M_{200}/\mbox{Gyr}$. The
analysis of the density profiles of cluster galaxies in different
redshift bins has allowed estimates of the {\em stellar} mass
accretion rates \cite{elli01} , $\dot{M_{\star}} \sim 0.06 \,
M_{\star}/\mbox{Gyr}$ at $z=0.45$ and $\dot{M_{\star}} \sim 0.02 \,
M_{\star}/\mbox{Gyr}$ at $z=0.20$, implying a significant
$z$-evolution of the mass accretion rate.  These estimates depend on
the timescale for halting star formation (SF in the following) in the
accreting blue galaxies, assumed here to be $\sim 1$~Gyr. A timescale
of $\sim 0.5$~Gyr would double these estimates, and make them
more similar to those obtained by the other techniques mentioned
above.

\subsection{Subclusters}
\label{subs}
In Sect.~\ref{accrete} evidence has been provided that clusters evolve
by accreting galaxies from the surrounding field.  A large fraction of
field galaxies occur in groups, hence in fact clusters not only
accrete isolated galaxies but also groups of galaxies.  When groups of
galaxies enter the cluster environment they are subject to tidal
forces that tend to disrupt them (see, e.g.,
\cite{gonz94,torm98,delu04}). The time of disruption is
longer for less massive groups \cite{torm98,delu04}, that can resist
for several Gyr \cite{torm98,prok07}.  Observationally, groups
accreted by clusters and not yet totally disrupted will appear as
secondary, statistically significant overdensities in the distribution
of cluster galaxies. They are usually called ``subclusters'' or
cluster ``substructures'' (see \cite{gira02b} for a review).

In reality, not all observationally-identified subclusters are the
surviving remnants of galaxy systems which have been accreted into a
cluster gravitational potentials.  Groups that are in the cluster
foreground or background can also be identified as local galaxy
overdensities, and hence confused with real subclusters. Among the
foreground and background groups, those that are dynamically bound to
the cluster will eventually be accreted and become real
subclusters in the future \cite{west90}.

\begin{figure}
\center{\includegraphics[height=9cm]{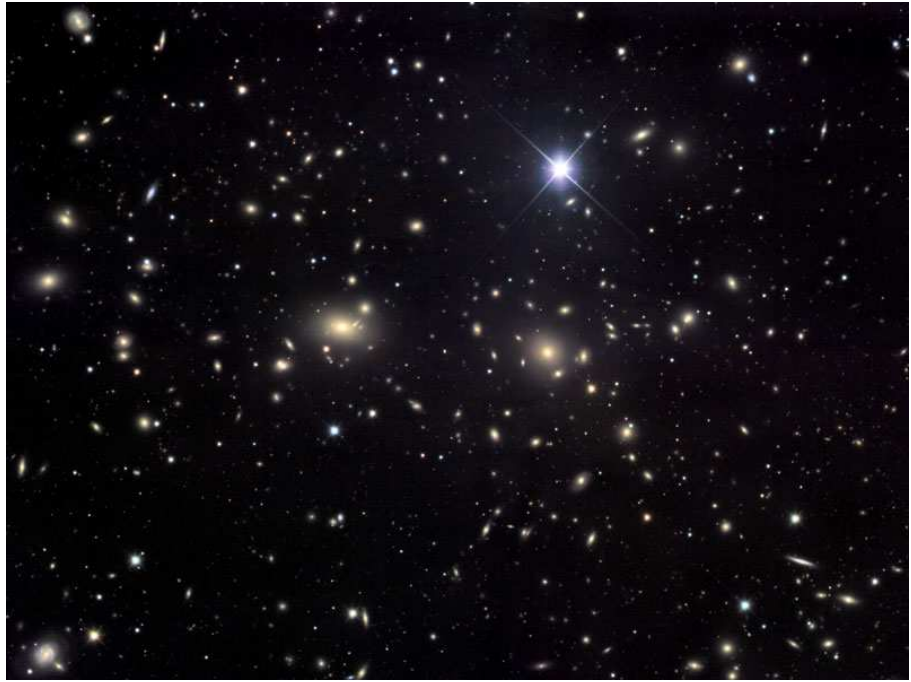}}
\caption{The central region of the Coma cluster ($1.13 \times 0.65$
  Mpc$^2$ at the cluster redshift, $z=0.0233$) displaying a clear
  bimodal distribution of galaxies concentrated around the two
  BCGs (from
  \texttt{http//www.mistisoftware.com/astronomy/Galaxies\_ComaCluster.htm)}.}
\label{fig-coma}
\end{figure}

It is important to identify and study subclusters. They provide
information on the accretion process itself and ultimately serve as
constraints to cosmological models (see, e.g., \cite{rich92,mohr95}).
For instance, the total mass of subclusters in the Coma cluster region
(see Fig.~\ref{fig-coma}) has been used to provide an estimate of the
Coma cluster accretion rate (see Sect.~\ref{accrete}). The
distribution of subcluster luminosities has been compared to the
distribution of subhalo masses thereby providing a test for theories
of structure formation \cite{rame07}. In the so-called 'bullet'
cluster the distributions of the total and baryonic masses are
displaced as the result of a very energetic collision of the cluster
with an infalling group \cite{barr02}. A measure of this displacement
has been used to set an upper limit to the cross-section of DM
particles \cite{mark04,clow06}.

Another aspect of the importance of subcluster studies is that
internal cluster dynamics can be affected by the very energetic ($\sim
10^{55}$~J) cluster--subcluster collisions. Part of the energy and of
the angular momentum of the collision is transferred to the cluster
and subcluster galaxies. As a consequence, the collision affects the
velocity distribution of cluster members, typically broadening,
skewing, and/or flattening an initially Gaussian distribution
\cite{bird93,pink96}, but also generating mean velocity gradients
along the collision axis \cite{cald98,lima97}.  Also the spatial
distribution of galaxies is affected, and becomes
less centrally concentrated \cite{pink96,bivi02}.  As a consequence, 
cluster masses and velocity dispersions are generally overestimated
during and some time after the collision
(see, e.g., \cite{bird95,pink96,bivi06}). Typically, masses are
overestimated by $\sim 30$\%, but depending on many parameters (e.g.
the relative angle of the collision and line-of-sight axes, the mass
ratio of the subcluster and main cluster, the time after the collision,
the observational sampling, etc.) the mass can be
overestimated by up to a factor $\sim 5$ or even {\em under}estimated
by a factor $\sim 3$ in extreme cases \cite{pink96,bivi06}.

Cluster--subcluster collisions can also have important effects on
galaxy properties.  They produce rapid variations in the cluster
gravitational field that can stimulate non-axisymmetric perturbations
in the galaxies involved in the collisions, and increase the rate of
galaxy--galaxy interactions, leading to bursts of SF
\cite{bekk99,gned03,cort06,kapf06}.  Moreover, collisions
are likely to displace the central BCG from the bottom of the cluster
potential well, thus effectively halting the accretion process of
satellite galaxies onto the BCG (see, e.g., \cite{gebh91,pimb06,rame07}).

Several methods have been developed to identify subclusters in the optical
(see, e.g., \cite{pink96} and \cite{gira02b}).
Depending on the data-set, the different
methods are more or less effective. Generally speaking, all these
methods look for deviations from symmetry in the spatial and/or
velocity distribution of cluster galaxies, or for significant secondary
peaks in the surface density or projected phase-space distributions.

The most widely used of these techniques has been developed in the
late 80s \cite{dres88}.
In its original formulation, the method consists in considering all
possible subgroups of 10 neighbors around each cluster galaxy. The
mean velocities and velocity dispersions of all these subgroups are
calculated, as well as their differences with respect to the
corresponding global cluster quantities. The sum of the squares of
these differences constitutes the $\Delta$ statistics,
\begin{equation}
\Delta = \sum_{i=1}^N (11/\sigma_v^2) [(\overline{v}_{i}-\overline{v})^2+(\sigma_i-\sigma_v)^2],
\end{equation}
where $\sigma_v$ and $\overline{v}$ are the velocity dispersion and
mean velocity of the whole cluster, $\sigma_i$ and
$\overline{v}_{i}$ are the corresponding quantities for any group $i$
of 11 galaxies, and the sum is over all $N$ cluster galaxies.
Montecarlo simulations are then run to
establish the statistical significance of $\Delta$. 

After its original formulation, this method has been modified and
adapted by several authors \cite{bird94,esca94,bivi02}.  This was done
in particular for extending the scope of the method, initially meant
to estimate the probability that a cluster contains subclusters, and
later adapted to find {\em which} galaxies have the highest likelihoods of
residing in subclusters.

Among clusters analyzed with the $\Delta$ technique, $\simeq 1/3$ show
significant evidence for subclustering, both at low and medium-to-high
redshifts (see, e.g., \cite{bivi97,milv08}). A similar fraction has
been found with other techniques (see, e.g., \cite{flin06,lope06}),
but this fraction is likely to be a lower limit, since other, deeper
analyses have discovered subclusters in clusters previously thought to
contain none (see, e.g., \cite{esca94,rame07}).  Presumably any
cluster would show evidence for subclustering if examined in
sufficient detail, because of the very nature of the process by which
these objects form (hierarchical clustering). Today's analyses are
aimed at determining not the {\em fraction} of clusters with
subclusters, but the {\em mass distribution of subclusters}
\cite{rame07} and its redshift evolution, and to compare it with the
prediction of cosmological numerical simulations, in order to learn
about the process of structure formation and evolution.

Other constraints can come from comparing the observed
distribution of subclusters in clusters with that
predicted by cosmological models. Observational estimates
of the number density profile of subcluster galaxies agree with
estimates from $\Lambda$CDM cosmological numerical simulations
\cite{bivi04,delu04}. Moreover, the subcluster orbits are found to be
tangential \cite{bivi04}, which is consistent with the idea that
subclusters on radially elongated orbits are selectively destroyed by
tidal effects \cite{tayl04,reed05b}.

\subsection{Summary and perspectives}
\label{dynsumm}
The analyses of cluster dynamics based on galaxies as tracers of
the gravitational potential have come to the following conclusions.
\begin{itemize}
\item $M(r)$ is consistent with the prediction of CDM cosmological
  numerical simulations.
\item A cored $M(r)$ is not excluded, but the core, if exists, has to
  be small, of order the size of the central bright galaxy,
  essentially ruling out self-interacting DM as a way to explain
  galaxy rotation curves.
\item At large clustercentric radii, $r \geq 2 r_{200}$, the mass
  density profile slope is $\simeq -3.5 \pm 0.5$, still consistent
  with NFW, but somewhat steeper.
\item The red/early-type/passively-evolving cluster galaxies are
  characterized by nearly isotropic orbits, while the
  blue/late-type/star-forming cluster galaxies have increasingly
  radial anisotropy with increasing clustercentric radius.
\end{itemize}

There is a lot that remains to be done and that will be made possible
by exploiting already existing and forthcoming databases, such as the
Imacs Cluster Building Survey\footnote{see
  \texttt{http://www.ociw.edu/research/adressler/}}. The constraints
that have been obtained so far can be put on a more solid statistical
basis. E.g. it should be possible to rule out either the cuspy NFW or
the cored Burkert with a $\sim 5$ times larger data-set than the ones
used so far. The currently loose constraints on the relation between
mass and concentration can be made tighter, in order to confirm or
reject the apparent (albeit marginal) discrepancy with the theoretical
predictions \cite{buot07,duff08}. While constraints on the {\em shape}
of $c=c(M)$ relation can only come by sampling the group mass scales,
constraints on the {\em normalization} can also come by sampling the
cluster mass scales alone, where kinematical methods are more powerful
(because the number of available galaxies per cluster is larger). On
the other hand, constraining the {\em group} $M(r)$ is extremely
important by itself, since group masses are intermediate between
cluster masses, where cosmologically motivated $M(r)$ models appear to
work, and galaxy masses, where they do not.

Individual cluster samples with $\geq 500$ galaxy velocities are
currently rare. With future, larger samples it will become possible to
constrain individual cluster concentrations and eventually compare the
$c$-distribution of a complete cluster sample with theoretical
predictions, which indicate a skewed distribution (e.g.
\cite{fede07}).  Maybe it will even be possible to check whether 
it is indeed the concentration that changes from cluster to cluster,
or whether it is the shape of the density
profile \cite{merr06,rico07}.

Models and simulations indicate that cluster mass profiles depend on
their accretion history \cite{zent05,hiot06,shaw06,hoff07,neto07}.
This could be tested by characterizing the mass profiles of galaxy
clusters as a function of their degree of internal dynamical
relaxation (or, inversely, of subclustering). Determining the detailed
properties of subclusters in a complete cluster sample can provide
another interesting test of cosmological models. Knowledge of the
cluster and subcluster masses, and their relative velocities
is required \cite{haya06}.

With a substantially larger data-base that the ones used so far, it
will also be possible to determine orbital constraints for several
classes of cluster galaxies, distinguished by color and internal
structure.  These are galaxy properties that are unlikely to evolve
simultaneously (see Sect.~\ref{colors}) and orbital isotropization can
be used as a clock for galaxy evolution in clusters.  It is also
important to determine the orbital characteristics of different
cluster galaxy populations as a function of cluster mass and/or
cluster structure, since orbital isotropization may occur at different
epochs and proceed with different speeds depending on the global
cluster properties. As a matter of fact, in some clusters
ETGs and LTGs move with similar orbital anisotropies \cite{hwan08}.

Numerical simulations and analytical models predict that the shapes of
the mass-density and anisotropy profiles of
cosmological halos are closely related \cite{hans06,barn07,falt07}.  
With current data-sets it is not possible to independently constrain cluster
mass- and anisotropy-profiles to the accuracy level required to test
this predictions, but with future, larger data-sets, it will.

Investigating the redshift {\em evolution} of cluster mass- and
anisotropy-profiles and of cluster mass accretion can provide
interesting constraints on the hierarchical model of cluster formation
and galaxy evolution in clusters. Cosmological numerical simulations
predict two phases of CDM halo assembly. The initial gravitational
collapse phase is responsible for establishing the central slope of
the mass density profile, while the external slope is established by
the later accretion phase \cite{lu06,lapi08}. It should then be
possible to test this evolutionary scenario by determining
the mass density profiles of clusters at different redshifts.

As far as the mass accretion rate is concerned, this is predicted to
increase with $z$ by cosmological simulations (e.g.  \cite{delu07}),
in a way that depends on the cosmological parameters $\sigma_8$ and
$\Omega_m$, and which can be observed as an increase of subclustering
or of cluster ellipticity (see, e.g., \cite{rich92,ho06}). The
evolution of cluster subclustering in different redshift bins can
therefore be used as a cosmological probe.  Features indicative of
massive subclustering seem to be found more often in distant clusters
than in nearby ones \cite{gira02b,dema06}, as expected. Although some
results on the ellipticity evolution with redshift have been obtained
from samples of nearby clusters \cite{plio02,lee06}, a more systematic
analysis of a well-controlled sample of clusters over a wide redshift
range is still lacking.

\section{Properties of cluster galaxy populations}
\begin{figure}
\center{\includegraphics[height=12cm,width=10cm]{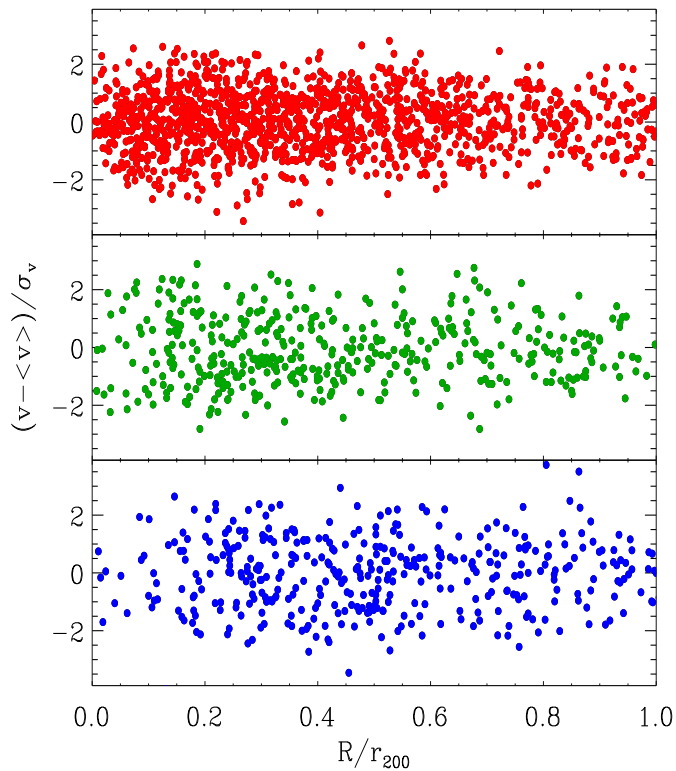}}
\caption{The projected phase-space distributions of different galaxy
  populations in ENACS clusters. Top panel: Es and S0s; middle panel:
  early-type S; bottom panel: late-type S and Irr. Galaxy
  clustercentric distances have been scaled with the cluster virial
  radii, and galaxy velocities relative to their cluster mean
  velocities have been scaled with the cluster velocity dispersions,
  in order to stack 59 cluster samples together. Galaxies in
  subclusters have been removed from the sample.  From \cite{bivi02}.}
\label{fig-rv}
\end{figure}

The most striking characteristics of the cluster galaxy population is
its morphology mix. Nearby clusters contain a larger fraction of ETGs
than of LTGs, exactly the opposite of what is observed in the general
field.  The morphological difference is so striking that it was noted
even before the extragalactic nature of {\em nebulae} was established
(see, e.g., \cite{bivi00}). Today, a modern version of the relation
between galaxy properties and their environments is expressed by the
segregation of different cluster galaxy populations in projected
phase-space (see Fig.~\ref{fig-rv}). ETGs populate
higher-density cluster regions than LTGs, which are also characterized
by a larger $\sigma_v$ at all radii. The very bright
ellipticals (Es) occupy denser and lower-velocity regions than any
other cluster galaxy population. Within the LTG population, the
projected phase-space distribution of early-type spirals (Sa--Sb) is
more similar to that of the ETGs than to that of late-type spirals and
irregulars (Irr).  Finally, galaxies in substructures appear to be
avoiding the cluster centers and to move at relatively low velocities.

Why do galaxies care about their environment? Why different
populations of galaxies occupy different regions of the phase-space?
By analyzing the properties of galaxies as a function of their
environment and of redshift it is possible to constrain the mechanisms
of galaxy formation and evolution (see,
e.g., \cite{balo00b}). Hereafter, the observational properties of
galaxies in and around clusters, as a function of their environment
and redshift are presented. Physical processes capable of shaping
galaxy properties in clusters are then reviewed.  Finally, an
evolutionary scenario based on the observational phenomenology is
proposed.

\subsection{Morphology}
\label{morph}
The galaxy morphology mix changes with some regularity over many
decades of projected galaxy number density, the fractions of Es and
S0s increase with increasing local density and the fractions of S and
Irr decrease \cite{dres80,post84}. This is the so-called
``morphology--density relation'' (MDR in the following). Given the
strong anti-correlation between local density and clustercentric
distance (radius), the MDR is often reported as a variation of the
morphological fractions with radius (see, e.g.,
\cite{whit93}).

\begin{figure}
\center{\includegraphics[height=11cm,width=11cm]{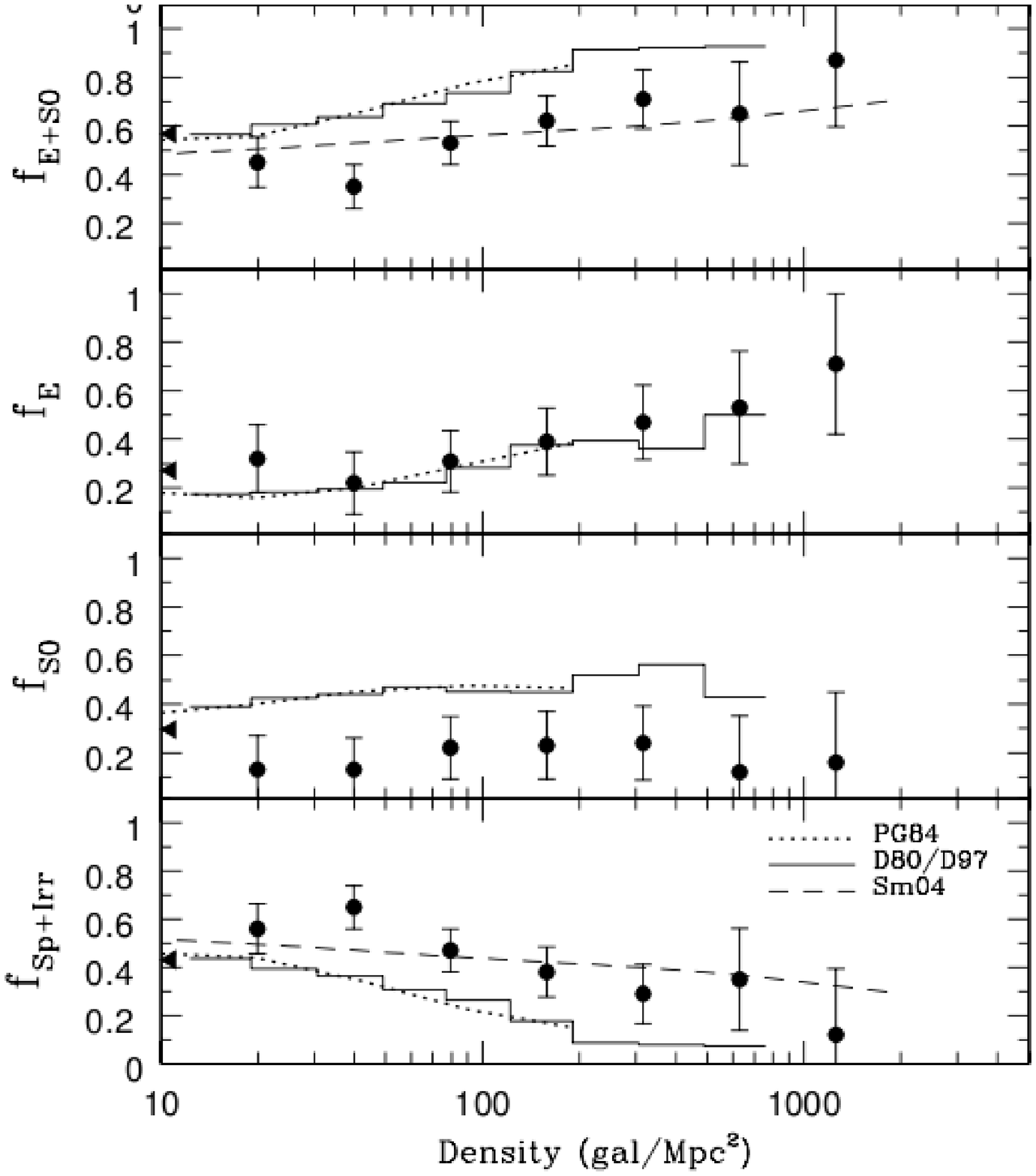}}
\caption{The MDR for $z \sim 1$ clusters (circles with error
  bars from \cite{post05}, dashed line from \cite{smit05}) compared
  with the same relation in local clusters (solid line from
  \cite{dres80,dres97}, dotted line from \cite{post84}). From
  \cite{post05}.}
\label{fig-mdr}
\end{figure}

A less strong correlation exists between the morphologies and the
velocities of cluster galaxies, Es and S0s having a narrower velocity
distribution than S and Irr (\cite{tamm72,moss77,sodr89,adam98c}; see
also \cite{bivi00} and references therein).  The MDR and the
morphology--velocity relation together can be viewed as a {\em
  segregation} of different cluster galaxy populations in projected
phase-space \cite{bivi02}.

The MDR is present at least up to $z \sim 1$ in massive clusters
\cite{post05,smit05}, while in low-mass, irregular
clusters at $z \sim 0.5$ there is no evidence of the
MDR \cite{dres97}. The MDR is anyway evolving also in massive
clusters, as the ETG fraction ($f_{ETG}$) decreases from $z \sim 0$ to
$z \sim 1$
\cite{dres97,fasa00,goto04,post05,smit05} (see Fig.~\ref{fig-mdr}).
The observed $f_{ETG}$ evolution does not concern Es, which seem to be
in place in the cluster environment well before $z>1$, but S0s, or
rotationally supported spheroidals \cite{mora07b}, whose fraction
increases with decreasing $z$ at the expense of S and Irr in the
highest density regions. The S0 fraction is $0.46 \pm 0.06$ in local
clusters, and decreases by a factor $\sim 2$ in $z \sim 1$ clusters
\cite{post05}.  In $z \sim 1$ clusters, S0s are $\sim 0.5$ Gyr younger
than Es of similar luminosities \cite{mei08}.  Most of the S0 fraction
evolution occurs at $z \leq 0.5$ \cite{desa07} and concerns
low-luminosity galaxies \cite{hold07}.

\subsection{Colors}
\label{colors}
Most cluster galaxies are distributed in a narrow band in a color vs.
magnitude diagram, the so-called color-magnitude relation,
CMR hereafter \cite{baum59,sand72,bowe92}. In nearby groups
and clusters the CMR is defined all the way down to dwarf
galaxies \cite{hilk03,carr06,penn08}. The CMR is well defined
also in high-$z$ clusters \cite{andr08,kurk08,mei08} and even for $z \approx
2$ protoclusters, but not so much for $z \approx 3$ protoclusters
\cite{koda07}. The mean color of the CMR galaxies indicates
their stellar populations were formed at $z \geq 2.5$
\cite{elli06,mei06a,delu07b}.

The CMR is rather tight in nearby clusters, but its mean color and
scatter increase with clustercentric distance and decreasing local
density \cite{abra96,pimb02,wake05,pimb06b} and at the fainter end of
the LF \cite{smai98}.  The increased scatter suggests a younger
average stellar population for the CMR galaxies in lower density
regions, with an age gradient of 2.5 Gyr from the cluster center to
its outskirts \cite{wake05}. A smaller age gradient is also seen in
high-$z$ clusters \cite{mei08}.  Spectral analyses show the field ETGs
to be $\sim 1$ Gyr older than their cluster counterparts \cite{bern06}.

\begin{figure}
\center{\includegraphics[height=9cm,width=11cm]{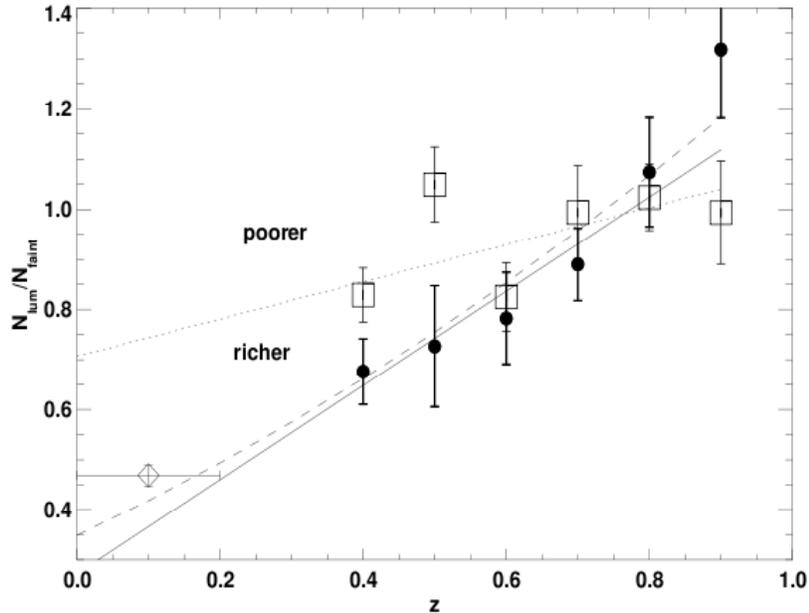}}
\caption{The ratio of luminous-to-faint CMR galaxies as a
  function of redshift, shown separately for richer (circles) and
  poorer (squares) clusters. A low-$z$ reference is also shown
  (diamond). From \cite{gilb08b}.}
\label{fig-cmr}
\end{figure}

The CMR faint-end becomes less populated in high-$z$ clusters
\cite{delu04b,tana05,delu07b,tana07} and this evolution occurs at different
epochs depending on the cluster richness
\cite{gilb08b} (see Fig.~\ref{fig-cmr}). Hence the faint-end of the red
cluster galaxy LF flattens \cite{toft04,muzz06}, and
the dwarf-to-giant red-galaxy number-ratio decreases
\cite{goto05,koya07,stot07}, with increasing $z$. As the number of
faint red cluster galaxies diminishes with $z$, the number of blue
cluster galaxies increases \cite{gilb08b}. Such variations of the
CMR with magnitude and $z$ are however not seen in all
clusters \cite{andr06,eise07,andr08}, and this indicates that there is
a significant cosmic variance in the evolution of cluster galaxy
properties \cite{wake05}.

The $z$-evolution of the CMR was already noticed 30 years ago by
Butcher \& Oemler \cite{butc78,butc84}. They measured the red-galaxy
fractions in two $z \simeq 0.4$ clusters, and found them to be
significantly lower than in nearby clusters. This phenomenology has
since been known as the Butcher-Oemler effect (BOe hereafter).  The
BOe can also be seen as an increase of the clustering strength of blue
galaxies with increasing $z$ \cite{mene06}. The BOe was initially
suspected of being caused by field galaxies contamination, but
spectroscopic observations have later confirmed the physical reality
of the effect. As the fraction of blue cluster galaxies increases with
$z$, also the fraction of galaxies with spectra characteristics of a
young stellar population (e.g. the E+A galaxies\footnote{E+A galaxies
  are galaxies characterized by an elliptical-like spectrum with
  strong Balmer lines \cite{dres83}.}) increases
\cite{dres99,pogg99,pogg06}, a phenomenology known as the {\em
  spectroscopic} BOe.  IR observations have discovered a population of
IR-bright emitters in medium-$z$ clusters, not seen in nearby clusters
\cite{fadd00,duc02,coia05,geac06,dres08,geac08,sain08}, a
phenomenology known as the IR-BOe.  The BOe is stronger for fainter
galaxies \cite{marg01,depr03} and in the more external parts of galaxy
clusters \cite{rako00,elli01}.

Altogether, studies of the CMR and BOe indicate that the color
evolution of cluster galaxies occurs {\em later} for {\em fainter}
galaxies, in {\em lower-density} regions, and in {\em less massive}
clusters. The analyses of the MDR find exactly the same trends for the
morphological evolution of cluster galaxies. Since there is a clear
correlation between galaxy morphology and color, the obvious question
is whether the morphology and color environmental dependences describe
in fact the same physical phenomenon. There is evidence that galaxies
in low-$z$ clusters do not show any MDR {\em at fixed age} (as deduced
from their spectral energy distributions), suggesting the MDR is in
fact an age--density relation \cite{wolf07}.  Similarly, the MDR of
galaxies in $0.4 \leq z \leq 0.8$ clusters can be derived from the
SF--density relation (which is strictly related to the age--density
relation) by using the average SF per morphological
class \cite{pogg08}. On the other hand, the radial increase of the CMR
scatter has been shown to be too strong to be entirely accounted for
by the MDR \cite{pimb02}.  Hence the relation between galaxy colors and
environment seems to be more fundamental than the MDR.

Morphological and color evolution do not seem however to proceed
at the same speed. Morphological evolution seems to take longer than
color evolution. This is indicated by the analysis of the evolution of
the galaxy mass function \cite{bund06}, by the presence of passive,
red cluster galaxies with spiral morphologies
\cite{goto03,pogg06,sanc07}, and by the presence of early-type S which
have the same age of S0s in the highest-density cluster regions
\cite{wolf07}.

\subsection{Masses}
\label{masses}
The masses of cluster ETGs can be inferred from the analysis of the
ETG fundamental plane (FP), a relation between the ETG internal
velocity dispersions, their effective radii, and their effective
brightnesses \cite{dres87,jorg96,mora07a}. By studying the FP as a
function of $z$, and based on spectrophotometric models, it is
possible to constrain the ETG $M/L$ evolution, their average formation
redshift ($z_f$), and the evolutionary history of their stellar
populations \cite{kels97}. FP studies indicate $z_f \geq 2$ for the
most massive cluster ETGs, but more recent $M/L$ evolution and younger
ages for ETGs of lower-masses \cite{kels97,dise06,jorg06,vdma07b}.
These studies also indicate that cluster ETGs are $\sim 0.4$ Gyr older
than field ETGs \cite{vdok07}, and that this age-difference increases
with galaxy mass \cite{dise06}. The $M/L$ of S0s and E+A galaxies are
different from the $M/L$ of Es \cite{barr06,kels97}, and this
indicates that the former have experienced more recent episodes of SF.

Overall, the results obtained from the study of the FP
support the conclusions reached from the studies of the CMR
and MDR.

\subsection{Luminosities}
\label{lums}
\begin{figure}
\center{\includegraphics[height=7cm,width=8cm]{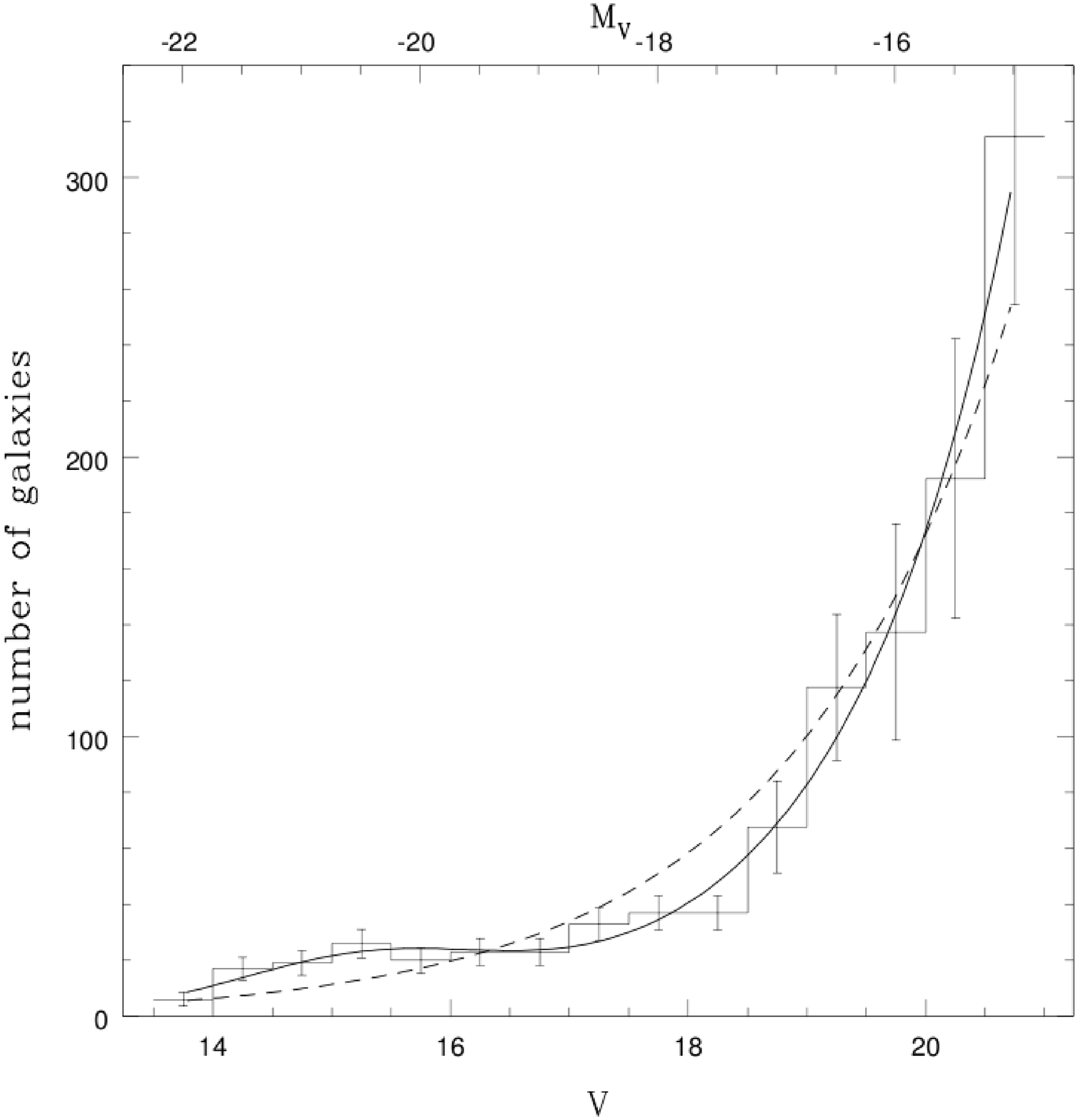}}
\caption{The LF of the Coma cluster of galaxies, showing the excess of
  bright galaxies relative to a Schechter function (dashed line), a
  steep faint-end upturn and a plateau at intermediate luminosities.
  The solid line is a Gaussian+Schechter fit to the data. From
  \cite{lobo97}.}
\label{fig-lf}
\end{figure}

Many analyses have shown the LF of group and cluster galaxies to be
different from the LF of field galaxies. In particular, compared to
the LF of field galaxies, the LF of cluster galaxies is characterized
by a steeper faint-end slope
\cite{driv94,depr95,lobo97,kamb00,cons02,durr02,chri03,depr03,saba05,tren05,pope06,adam07,jenk07,miln07,yama07,boue08},
a brighter characteristic magnitude
\cite{mart02,mahd05,robo06,zand06}, and a dip or plateau at
intermediate luminosities \cite{thom93,driv94,bivi95,mile06,pope06}
(see Fig.~\ref{fig-lf}).  According to other studies, however, there
is no significant difference between the LF of cluster and field
galaxies \cite{paol01,tren01,tren02,lin04,hars07,penn08,rine08b}.

Different conclusions about the shape of the average LF of cluster
galaxies may be reached if different clusters are characterized by
different LFs, or if different studies consider different cluster
regions and the LF changes with radius. While the cluster LF appears
to be universal (see, e.g., \cite{pope06}), there is evidence that its
shape does change with radius, its faint-end slope being shallower
(and hence similar to the field LF) in the central regions of clusters
\cite{lobo97,kamb00,andr01,durr02,odel02,saba03,hain04,prac04,saba05,bai06,pope06,boue08}.
It is also possible that the reported faint-end slopes of cluster LFs
are overestimated because of incorrect field-count subtractions
\cite{valo01,hilk03,boue08}, or, vice versa, underestimated because of
a biased selection against low surface-brightness galaxies
\cite{driv04}.

The ratio between the number of bright (giant) and of faint (dwarf)
galaxies increases towards the cluster center, an effect known as {\em
  luminosity segregation} \cite{bivi00,cape81}. Very bright cluster
galaxies not only prefer the central regions, but are also closer to
the mean cluster velocity, their $\sigma_v$ being only $\leq
1/2$ the global cluster $\sigma_v$ \cite{bivi92,adam98c}.
Luminosity segregation is also found outside clusters (e.g.
\cite{norb02,gira03}).

\begin{figure}
\center{\includegraphics[height=8cm,width=10cm]{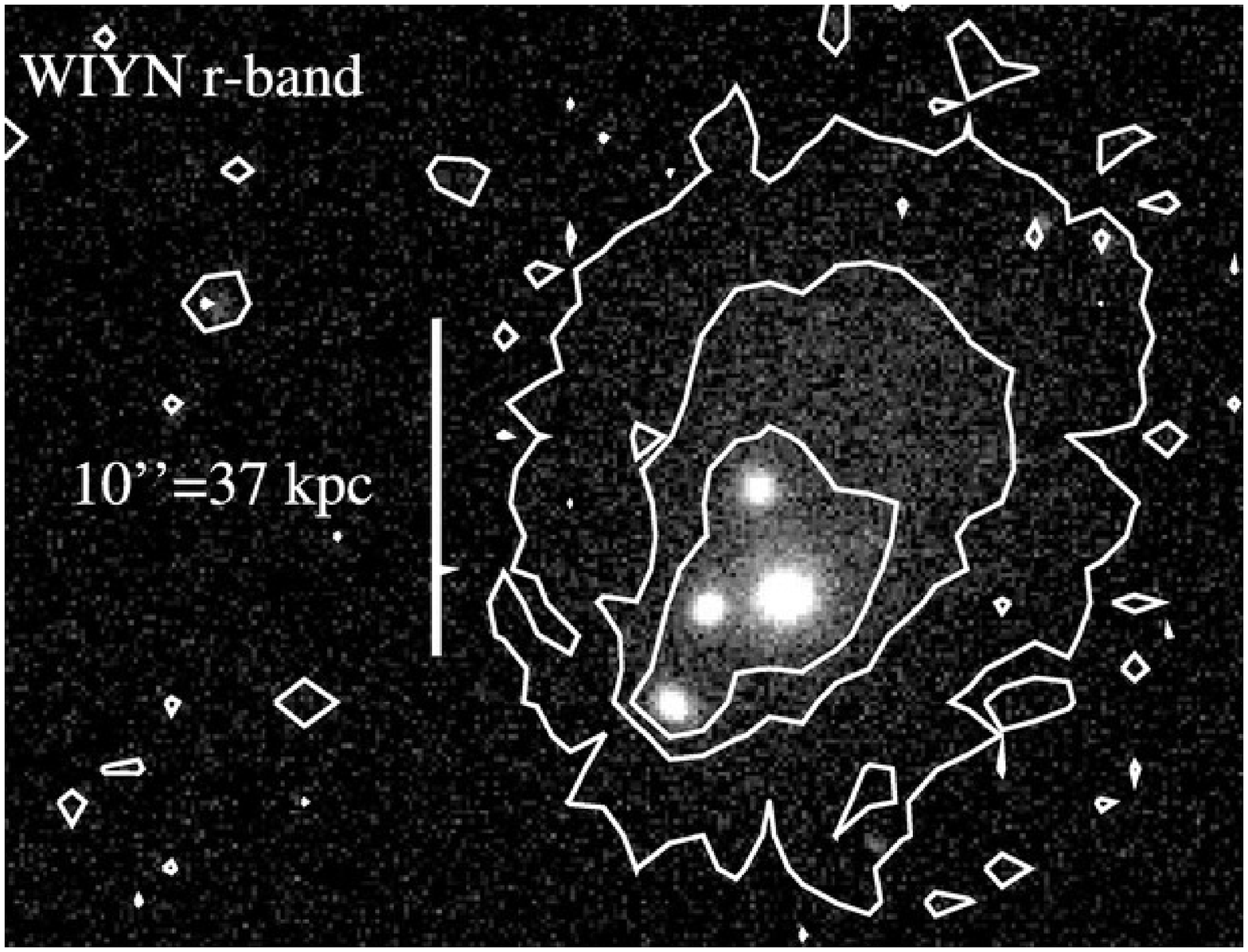}}
\caption{The central region of a $z=0.39$ X-ray selected galaxy
  cluster.  Three companion galaxies are seen close to the
  BCG. Contours show $r$-band surface brightness levels and
  emphasize the presence of the ICL. Given the positions and
  velocities of the four galaxies, two of the companions are estimated
  to merge with the BCG in $\sim 0.1$ Gyr. From
  \cite{rine07}.}
\label{fig-bcg}
\end{figure}

Two other characteristics distinguish the luminosity content of
clusters from the field, the presence of BCGs
and of the intra-cluster diffuse light (ICL in the following). 
BCGs are quite different from field ETGs of the same
luminosities, in particular since their sizes are bigger
\cite{bern07,vdli07}.  Moreover, BCGs are intimately related
to their hosting cluster. In fact they live close to the bottom of the
cluster potential \cite{oege01,rame07}, their luminosities are
correlated with the cluster total luminosities and masses
\cite{lin04b}, and their main axes are aligned along the main cluster
axes \cite{stru87}.  

There is evidence of some ongoing SF activity in many BCGs, probably
activated by the cooling of the intra-cluster gas
\cite{egam06,edwa07}, but overall the average color of the BCG stellar
populations suggests they formed at $z \geq 2$ \cite{stot08}, and the
total $K$-band luminosity of massive cluster BCGs evolves passively
with $z$. On the other hand, BCGs in clusters of low masses do show
substantial stellar mass evolution, with a factor 2--4 increase in
mass since $z \sim 1$ \cite{brou02,nels02}.  This is consistent with
the fact that the fractional cluster luminosity in the BCG is
anti-correlated with the total cluster light \cite{lin04b}, as if BCGs
in massive clusters completed their assembly before their counterparts
in low-mass clusters.

\begin{figure}
\center{\includegraphics[height=7cm,width=8cm]{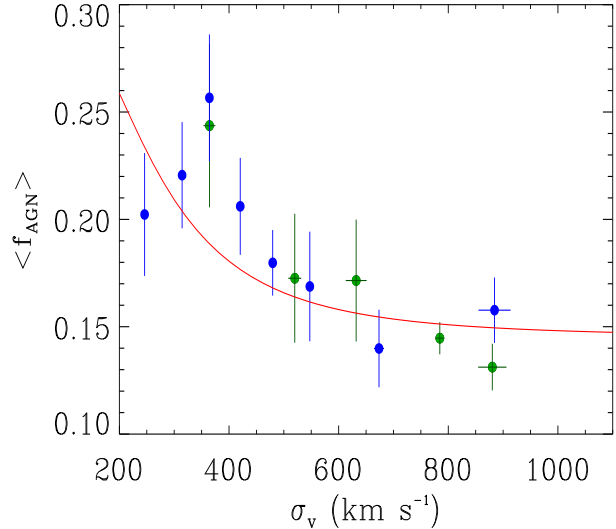}}
\caption{The AGN fraction as a function of the galaxy system velocity
  dispersion for two samples. The solid line is a model based on a
  galaxy merging-rate relation \cite{mamo92}. From \cite{pope06b}.}
\label{fig-fagn}
\end{figure}

Support for the hypothesis that at least some cluster BCGs
have yet to complete their assembly comes from the observation that the
luminosity ratio between the BCG and the second-ranked
cluster galaxy is smaller for the less massive and irregular clusters
\cite{loh06,rame07}.  Direct evidence for ongoing build-up of
BCGs in some $z \simeq 0.4$ clusters and groups has also been
provided \cite{rine07,tran08} (see Fig.~\ref{fig-bcg}).

The amount of ICL in groups and clusters is not very well constrained.
It is estimated to be $\sim 5$--25\% of the total cluster light
\cite{feld04a,feld04b,ague07,kri06,kri07,daro08}, and as much as 50\%
near cluster centers \cite{zibe08}.  The properties of the BCGs
and the ICL are related, suggesting a direct, physical link. Their
colors and elongation axes are in fact similar, and clusters with
brighter BCGs have higher ICL surface brightnesses \cite{zibe05}.

\subsection{Nuclear activity}
\label{agn}
There is observational evidence that the fraction of active galactic
nuclei ($f_{AGN}$) is higher in galaxy pairs \cite{wood07} and in
compact groups \cite{cozi00}, and lower in clusters \cite{dres99},
relative to isolated field galaxies.  However, other studies have
found no environmental dependence of $f_{AGN}$ \cite{mill03,elli08}.

The striking difference between $f_{AGN}$ in compact groups and in
clusters suggests it is not the density of the environment that
matters for the onset of the AGN activity. In fact, galaxy number
densities in compact groups and clusters are both very high. On the
other hand, cluster $\sigma_v$s are typically $\sim 3$ times
larger than those of compact groups. Hence, the onset of AGN activity
may depend more on the relative galaxy velocities, than on galaxy
density.  A study of several hundred groups and clusters has indeed
shown that $f_{AGN}$ is anti-correlated with the velocity dispersion
of the host galaxy system \cite{pope06b} (see Fig.~\ref{fig-fagn}).

\subsection{Physical processes}
\label{processes}
Several physical processes are capable of affecting the properties of
cluster (and group) galaxies. They result from the interactions among
galaxies, with the cluster gravitational potential, and with the
intra-cluster medium \cite{treu03}. The most important of these
processes are the following:
\begin{itemize}
\item dynamical friction \cite{chan43,boyl08};
\item galaxy-galaxy collisions \cite{spit51,moor96}, leading to tidal
  effects and, eventually, mergers \cite{negr83,barn92};
\item tidal forces induced by the cluster gravitational potential,
  resulting in tidal truncation of the galactic halos
  \cite{rich76,merr84,ghig98};
\item ram-pressure stripping of the galaxy gas
  \cite{gunn72,quil00}.
\end{itemize}

Dynamical friction is the process that slows down a massive galaxy as
it moves in a sea of dark matter particles.  These particles are
gravitationally focused in the wake of the galaxy itself. The galaxy
therefore feels a braking force resulting from the excess mass density
of the dark matter particles in the direction opposite to its
motion. The characteristic timescale of this process is
\begin{equation}
t_{df} \propto \frac{v_g^3}{m_g \rho},
\end{equation}
where $v_g$ is the galaxy velocity, $m_g$ its mass, and $\rho$ is the 
dark matter 
density \cite{chan43,esqu07}. Hence dynamical friction is more effective
in higher density environments and for more massive galaxies and not
very effective when the galaxy moves fast. 

Collisions and mergers drive substantial galaxy morphological
modifications \cite{barn92,moor99b}, and induce starburst episodes in
the central galaxy regions \cite{miho94,stru97,fuji98}, which then
consume the available gas content. Gas can also be expelled from the
galaxies as these collide and eventually merge \cite{barn92,duc08},
since the total mass of the merger product is generally less than the
sum of the progenitor masses. The characteristic timescale for a
galaxy to experience a collision with another galaxy is
\begin{equation}
t_c \propto \frac{1}{\nu r_g^2 v_g},
\end{equation}
where $v_g$ is the relative velocity between the two galaxies, $r_g$ 
the galaxy radius, and $\nu$ is the galaxy number density \cite{gned03}.
Hence collisions happen more frequently in higher density
environments and between larger galaxies.

The impulsive tidal field produced by the collision produces
collisional stripping \cite{merr84}. This is most important for the
material in the external regions of non-compact galaxies undergoing
close encounters, since the tidal force is proportional to $r_g/d^3$,
where $d$ is the distance to the colliding galaxy.

Galaxy-galaxy collisions may lead to mergers when the relative speed
of the encounter is very low. The characteristic timescale for a merger
to occur is
\begin{equation}
t_m \propto \frac{\sigma_v^3}{\nu r_g^2 \sigma_g^4},
\label{eq-merger}
\end{equation}
where $\sigma_v$ is the cluster velocity dispersion and $\sigma_g$ is
the internal velocity dispersion of the galaxy \cite{mamo92,maki97}.
Mergers are therefore very effective in low-velocity dispersion,
high-density environments, such as groups and cluster outskirts
(filaments).

A galaxy moving on a circular orbit in the core of a cluster with core
radius $r_c$ will suffer tidal truncation of its external parts, beyond
a radius \cite{merr84}
\begin{equation}
r_t \approx r_c \frac{\sigma_g}{2 \sigma_v}.
\label{eq-trunc}
\end{equation}
Hence galaxies moving in massive and centrally concentrated clusters
will be severely truncated, and their external (dark matter and
gaseous) halos will become part of the intra-cluster matter as they
come close to the cluster center \cite{ghig98}.  Note however that
galaxies sitting at the bottom of the cluster potential, i.e. BCGs,
are not truncated by this mechanism since they feel symmetric external
forces. Eq.~(\ref{eq-trunc}) may however not be applicable to
field galaxies entering the cluster along filaments, since their
orbits are mildly radially anisotropic, and not circular
\cite{ghig98,bivi04}. For these more general orbits,
numerical simulations indicate that eq.~(\ref{eq-trunc}) is still
valid, once $r_c/2$ is replaced with the radius of the orbit
pericenter \cite{ghig98,mamo00}.

The tidal forces induced by the cluster potential also lead to the
compression of galactic gas and can therefore stimulate SF
\cite{byrd90,fuji98}.

\begin{figure}
\center{\includegraphics[height=8cm,width=9cm]{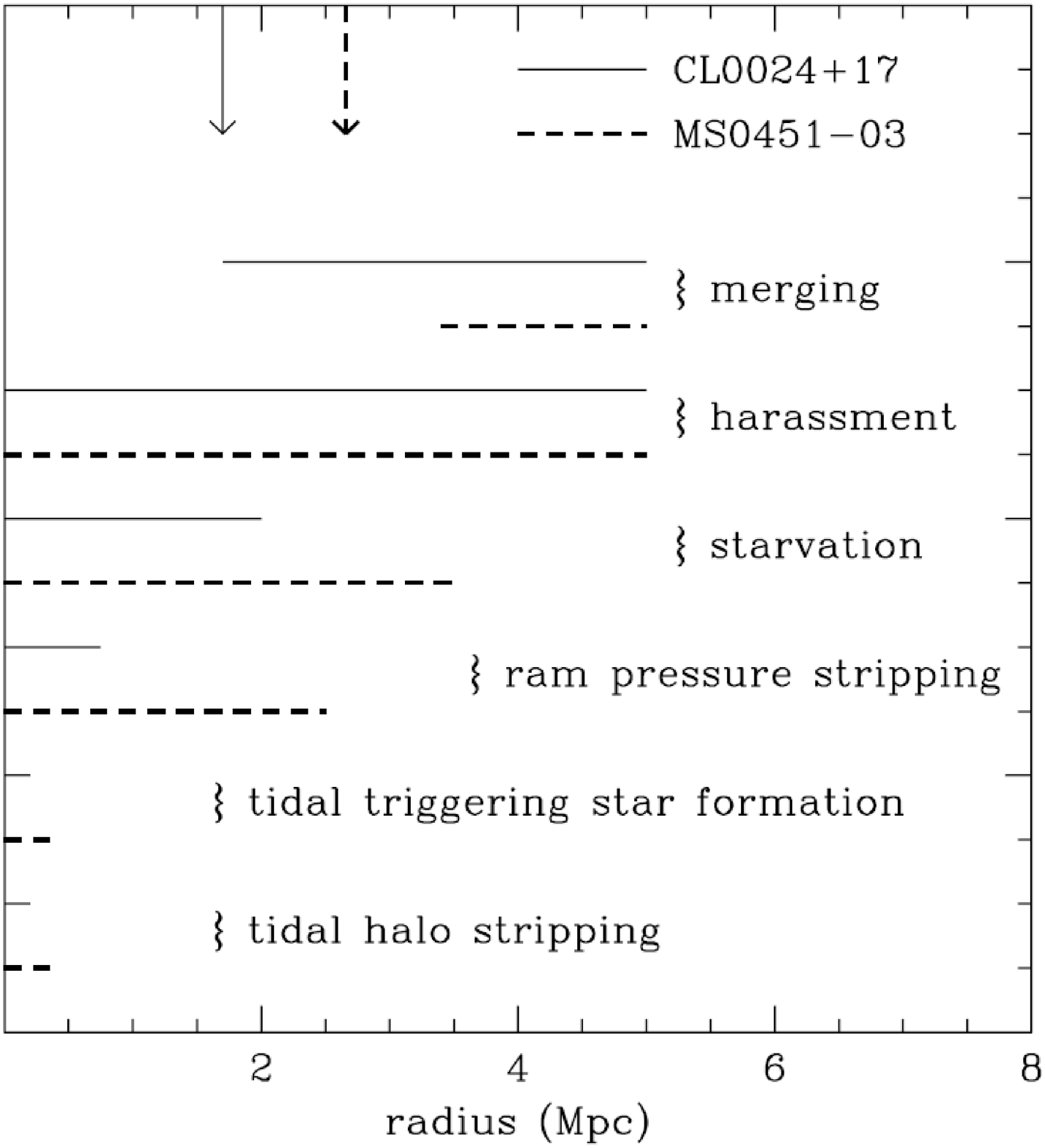}}
\caption{This diagram indicates the clustercentric radius over which
the listed mechanisms are effective in shaping galaxy properties. Two
clusters are considered, with $M_{200}$ masses $\simeq 9$ and $\simeq
14$ in units of $10^{14} M_{\odot}$ (Cl0024+17 and MS0451-03,
respectively). Arrows indicate the cluster virial radii. The indicated
tidal processes refer to interactions with the cluster potential,
while the term ``harassment'' refers to the effect of galaxy-galaxy
collisions. From \cite{mora07a}.}
\label{fig-moran}
\end{figure}

A galaxy moving in a cluster will feel a pressure exerted by the intra-cluster
diffuse gas onto the gas of its disk and the gas reservoir in its halo.
The disk gas will be stripped under the condition that
the ram pressure of the intra-cluster gas exceeds the gravitational
restoring force per unit area on the disk,
\begin{equation}
\rho_{IC} v_g^2 > 2 \pi G \Sigma_{\star} \Sigma_{gas},
\label{eq-ram}
\end{equation}
where $\rho_{IC}$ is the intra-cluster gas density, and
$\Sigma_{\star}$ and $\Sigma_{gas}$ are the stellar and gaseous
surface densities of the disk, respectively \cite{gunn72,mcca08}.  In
fact, eq.~(\ref{eq-ram}) is strictly valid only if the baryons dominate the
mass budget of the galaxy disk. As far as the gas in the galaxy {\em halo}
is concerned, the stripping condition can be written as follows
\begin{equation}
\rho_{IC} v_g^2 > \alpha \frac{G m_g \rho_{gas}}{R_g},
\label{eq-ram2}
\end{equation}
where $R_g$ is the projected galaxy radius in the direction
transverse to the galaxy motion, $\rho_{gas}$
is the 3-d galaxy gas-density profile and the $\alpha$ term,
of order unity, depends on
the precise shape of the gas and mass profiles of the galaxy
\cite{mcca08}.  From eqs.~(\ref{eq-ram}) and
(\ref{eq-ram2}) it is clear that it is easier to strip the gas from a
lower-mass galaxy moving at higher speed in a denser intra-cluster
medium.

Ram-pressure, tidal truncation, collisional stripping can all lead to
{\em starvation} \cite{lars80,balo00}. Present-day spirals would use
their disk gas content in a few Gyr at their current SF rates. If
these SF rates must be sustained over a Hubble time, the disks must be
refueled by gas reservoirs in galaxy halos.  Starvation results from
removing the gas reservoir.

It is important to note that the effectiveness of the
different processes depend on the clustercentric distances and
on the cluster masses. E.g. merging will operate more effectively in the
cluster outskirts and for lower-mass clusters, and ram-pressure
stripping in the cluster inner regions and for higher-mass clusters
(see Fig.~\ref{fig-moran}).

\begin{figure}
\center{\includegraphics[height=12cm,width=10cm]{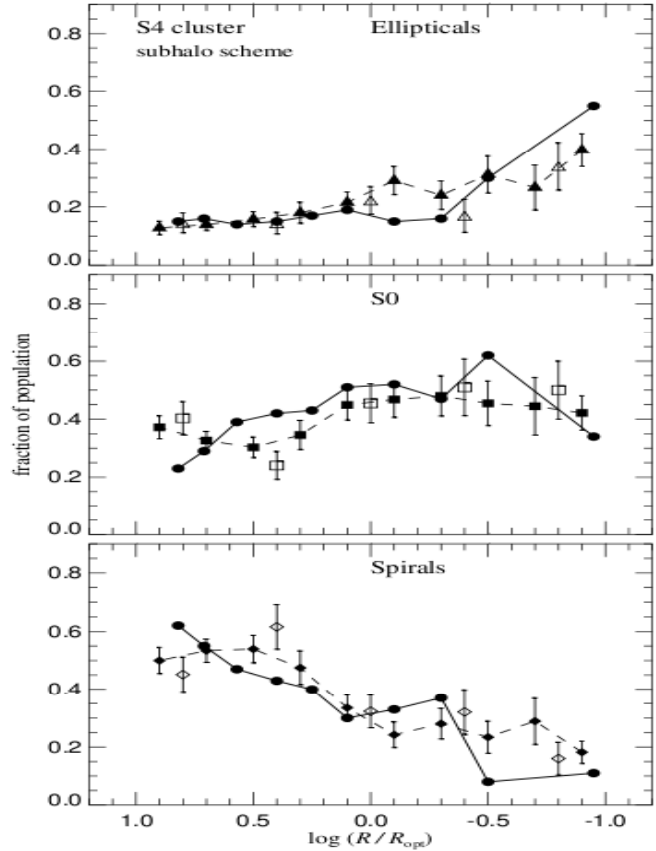}}
\caption{The morphological mix of galaxies as a function of
  clustercentric radius in simulations (dashed line; filled and open
  squares correspond to bright and faint galaxies) as compared to real
  data (solid line, from \cite{whit93}). The agreement is excellent
  but the definition of 'numerical S0s' has been widened to include
  'numerical early-spirals', otherwise a deficit of S0s compared to the
  observational sample would have resulted. From \cite{spri01}.}
\label{fig-mdrsim}
\end{figure}

\subsection{A scenario for the evolution of galaxies in clusters}
\label{scenario}
What makes the properties of cluster galaxies differ from those of
field galaxies? Do galaxies change their colors, morphologies,
luminosities, and even nuclear activities, as they move from regions
of lower density to regions of higher density under the influence of
gravity?  Or are galaxy properties established once and for ever since
their formation? The two scenarios confront each other, but are not
necessarily mutually exclusive.

Several characteristics of the cluster galaxy population argue in
favor of the scenario in which galaxy properties are established {\em
ab initio.} In particular, both the CMR and the MDR seem to be
established already in $z
\geq 1$ clusters, and the color and FP-inferred
$M/L$-evolution of bright cluster ETGs indicate simple passive
evolution since $z_f \geq 2$. However, other observations argue for a
modification of the galaxy properties with time. In particular, there
is a deficit of both faint ETGs and S0s in high-$z$ clusters, as
indicated by studies of the CMR, the MDR and the LF. This deficit is
more conspicuous for less massive and more irregular clusters.  The
analyses of the CMR, the FP, and the ETG spectra also show that the
average age of the ETG stellar population increases with the local
density.

A natural expectation of hierarchical cosmological models is {\em
biased galaxy formation} (e.g. \cite{cen93}), galaxies in denser
environments form earlier, and are more massive than those formed
later, because they form with an initially larger mass and undergo
more mergers during their lifetime.  These models are able to
approximately reproduce the environmental dependences of galaxy
properties
\cite{diaf01,spri01,avil05,lanz05,saro06} (see Fig.~\ref{fig-mdrsim}).
Mergers increase the masses of galaxies, destroy spiral features and
disks, reduce angular momentum.  They also trigger a starburst that
consumes part of the available galaxy gas. Mergers are therefore
effective in driving a LTG to ETG transition.

A merger can also form and/or activate a central black hole, hence an
AGN, perhaps with a certain delay after the
starburst \cite{li08}. This is observationally supported by the fact
that AGN host galaxies have colors and morphologies suggestive of a
1--4 Gyr old starburst associated with a merger event \cite{silv08},
and by the fact that a merger model fits the relation between
$f_{AGN}$ and $\sigma_v$ of galaxy systems \cite{pope06b}.
The resulting AGN feedback could quench any remaining SF activity
\cite{silk98,scan05}, more strongly so in more massive galaxies,
characterized by larger black hole masses \cite{spri05}. The AGN
energy output into the intra-group medium can also explain the
phenomenology of group X-ray scaling relations \cite{lapi05}.

\begin{figure}
\center{\includegraphics[height=7cm]{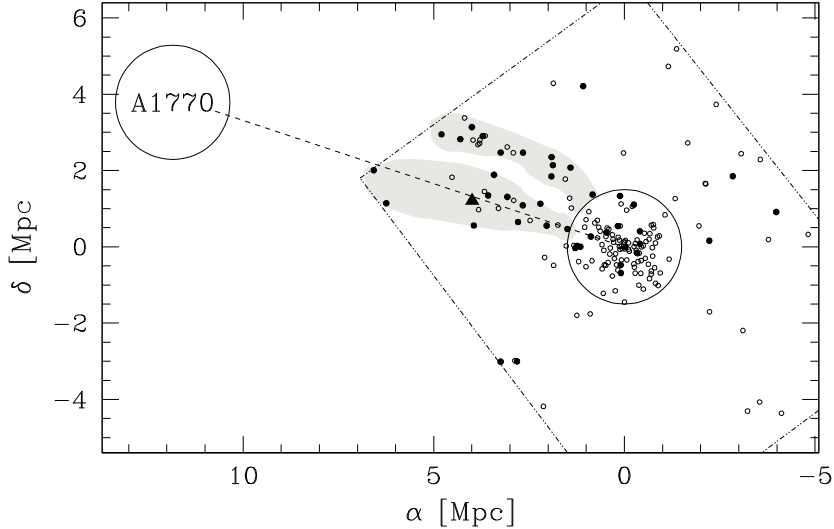}}
\caption{Spatial distribution of members of the A1763 galaxy cluster
at $z \sim 0.2$. Starburst galaxies identified by their IR emission
are indicated with filled dots. The triangle indicates an AGN. Circles
of 1.5 Mpc are drawn around the central positions of A1763 and its
neighbor A1770. Shaded regions highlight the kinematically-detected
filaments, where most starburst galaxies are located \cite{fadd08}.}
\label{fig-fila}
\end{figure}

Mergers of small galaxies or galaxy subunits should be able to form
red Es at the cluster centers at times when cluster $\sigma_v$s
have not yet become too high. A merger origin of cluster
Es is indeed suggested by their flat metallicity and color
gradients \cite{mehl03,laba04}. Mergers become inefficient at later
times in high density regions, when these become characterized by high
values of $\sigma_v$ (see eq.~\ref{eq-merger}).  But mergers
can still operate in lower-density regions. Hence, low-density regions
should be populated by galaxies forming at later times, and hence
grown up to smaller masses, in agreement with the observed trends of
galaxy age with density and radius inferred from the analyses of the
CMR relation. At intermediate $z$, the IR-BOe
is suggestive of mergers among galaxies located in medium-density environments
but outside the highest-density cluster regions
\cite{koya08}.  At lower $z$, the merger activity is restricted to
galaxies outside the cluster virial regions, as indicated by the high
fraction of starburst galaxies detected in the relatively
high-density, but low-$\sigma_v$ cluster-feeding filaments
\cite{brag07,port07,fadd08} (see Fig.~\ref{fig-fila}). Most of these starburst
galaxies have small stellar masses.

Summing up, high-density regions are the birthplace of massive
galaxies, which assembly most of their mass at early times, before
mergers become inefficient. Mergers stimulate SF and AGNs that
successively quench any further SF. In lower-density environments less
massive galaxies are formed, and the formation process is
delayed. SF hence stops earlier in more massive galaxies in
higher-density environments. At least qualitatively, this
scenario is consistent with the observation that the critical mass
above which SF is suppressed increases with $z$, and more rapidly in
denser environments \cite{nuij05,bund06}, the so-called {\em
``downsizing''} effect \cite{cowi96}.

According to numerical simulations, mergers of cluster galaxies can
also make BCGs \cite{dubi98,delu07}. Support for a merger origin of
BCGs comes from their shallow surface-brightness
profiles \cite{brou05}. BCG assembly can continue until very late, and
is only $\sim 50$\% finished by $z \sim 0.5$ \cite{delu07}.  Direct
observational evidence that BCGs continue their mass growth at $z<0.5$
has recently been provided \cite{rine07,tran08}.  The more recent mass
evolution of BCGs in lower-mass clusters \cite{brou02,nels02} is just
another manifestation of the {\em downsizing} process.  Cluster
$\sigma_v$s would generally be too high for BCGs to continue
growing at recent times, even in relatively low-mass clusters, but
dynamical friction can decrease the relative velocities of the more
massive cluster galaxies orbiting the cluster central
regions. Velocity segregation of the most luminous cluster galaxies is
indeed observed \cite{bivi92} and is also predicted by numerical
simulations \cite{spri01}. Dynamical friction and mergers of the
brightest galaxies in the central regions of galaxy systems can also
account for the intermediate-luminosity dip and the brighter
characteristic luminosity of the cluster LF, relative to the field
LF \cite{cava92,cava97,gonz05}.

Galaxy mergers must also be important in the formation of the ICL. The
ICL characteristics are clearly related to the characteristics of the
BCG, suggesting their common origin.  The product of galaxy mergers is
unlikely to contain 100\% of the mass of their progenitors, as part of
the stars of the initial galaxies are scattered away during the
galaxy-galaxy collision.  If $\sim 1/3$ of the stars of the merging
galaxies are scattered into the ICL component at each merger event, it
is possible to reconcile current hierarchical cosmological models with
the lack of evolution since $z \sim 1$ of the stellar mass function at
the high-mass end \cite{mona06,stra06,conr07}. Numerical simulations
predict that if the ICL is built through galaxy-galaxy collisions its
fractional contribution to the total cluster light should be almost
independent on cluster mass \cite{mura07}, a prediction consistent
with some observational estimates \cite{kric07}, and in contrast with
others \cite{gonz07}.  Anyway, direct evidence for the ICL creation
through galaxy-galaxy collisions comes from observations of compact
groups \cite{durb08}. Dwarf galaxies can form in the tidal tails of
galaxy-galaxy collisions \cite{barn00} contributing to steepen the LF
faint-end of galaxy clusters. These tidal dwarfs are then rapidly
destroyed in high-density regions by collisions with other galaxies,
thus explaining why a flatter LF characterizes cluster central
regions.

Mergers and dynamical friction are therefore important processes in
the evolution of cluster and group galaxies. On the other hand, the
abundance of S0 galaxies in clusters and its late-evolution are more
difficult to explain \cite{diaf01,spri01,avil05}. 
{\em Low-luminosity} S0s could form from spirals through the fading of the
spiral disks resulting from gas removal (starvation)
\cite{pogg01a,mora07a}.  The production of S0s from spirals is
suggested by the relative evolution of the LFs of blue and red galaxies
\cite{gilb08b} and by the presence of a cluster galaxy
population intermediate between ETGs and LTGs, the population of
passively-evolving galaxies with spiral arms and red disks
\cite{mora06,mora07a}. These {\em passive spirals} are 
generally also characterized by a very low-HI gas content given their
stellar mass or luminosity (see, e.g., \cite{hayn84,sola01}). The formation
of S0s via the fading of spiral disks is also suggested by the
analyses of the Tully-Fisher relation and the globular cluster
fraction of cluster S0s \cite{hinz01,bedr06,barr07}, as well as by the
smaller disk sizes of cluster spirals relative to field
spirals \cite{ague04,guti04}.

The precise mechanism by which cluster spirals loose their gas can
vary, both with the mass of the accreting cluster, and with the
distance from the cluster center \cite{domi01,mora07a}. In the central
regions of massive clusters, ram-pressure stripping is expected to be
the dominant mechanism \cite{voll01}, but tidal shocks in the cluster
potential can also be important if the cluster mass distribution is
highly concentrated \cite{moss00}. At larger radii, and in low-mass
clusters and groups, galaxy-galaxy collisions and related tidal
stripping should dominate \cite{balo00b,hels03,balo04}. These
collisions can produce central bursts of SF, accelerating the disk gas
consumption.  There is indeed conspicuous observational evidence that
at least some cluster (or group) S0s have experienced recent episodes
of SF
\cite{naka01,pogg01a,blak03,mei06a,tana06,mora07a,roge07,tran07}.
If processes operating in the cluster outskirts are effective
enough in removing the gas of infalling field spirals, 
processes operating near the cluster center may play
little r\^ole in the evolution of cluster galaxies
\cite{balo00b,okam03,lanz05}.

The fading of spiral disks reduces galaxy luminosities by $\geq 1$ mag,
hence it is impossible to form bright S0s by this mechanism. In order
to produce bright S0s from spirals, the spiral bulge mass must
increase.  This can occur by a central burst of SF
\cite{okam03} induced by minor mergers \cite{koda01,bour05} and/or major
tidal heating. The tidal heating can be caused by galaxy-galaxy
collisions or by the gradient of the cluster gravitational potential
\cite{gned03}. 
When galaxies collide at high speed, no merger can occur and tidal
damage is limited, hence bright cluster S0s can only form at high-$z$
before the cluster $\sigma_v$ becomes too large. At low-$z$,
bright S0s can form in small, low-$\sigma_v$ groups. As these
groups are accreted by clusters, they then become part of the cluster
galaxy population
\cite{cava97,treu03,kaut08}. Accreted groups are identified as
subclusters, and an excess of S0s in subclusters in the central regions
of nearby clusters has indeed been observed \cite{bivi02}.

Possible progenitors of the bright S0s are early-type spirals (Se). In
fact, the bulge luminosities of Se are comparable to
those of S0s \cite{thom02}, Se and S0s occupy regions
of similar (albeit somewhat lower) local densities
\cite{thom06}, the stellar populations of the Se and S0s that reside
in high density environments have similar ages \cite{wolf07}, and Se
and S0s move on similar, nearly-isotropic, orbits \cite{bivi04}.
Other plausible progenitors of the bright S0s are the E+A galaxies.
Spectrophotometric modeling indicates that E+A cluster galaxies
should passively evolve into bright S0s \cite{shio04}. E+A galaxies
have positive color gradients \cite{yama05} and have close companions
and/or tidal features indicative of interactions \cite{lave88}, all
features consistent with a merger origin of these galaxies. Recently,
a population of dusty starburst galaxies, bright enough to evolve into
bright S0s after the SF episode has ceased, has been identified in
$z \sim 0.5$ clusters, using observations with the {\em Spitzer}
satellite \cite{dres08,geac08}. Many of these presumed S0-progenitors
are found to reside in subclusters, where the galaxies can collide
at relatively low speeds.

Albeit very sketchy, the scenario outlined above tries to account for
a large part (if not all) of the observational evidences of the
environmental dependence of galaxy properties as a function of
$z$. With a precise theory of galaxy formation and evolution yet to be
formulated, and an infinity of data yet to be collected, our current
(rather modest) understanding of galaxy evolution in clusters will
certainly have to be revised in the near future.

\acknowledgments These lectures were partly written during a 
three-months ``Poste Rouge'' stay at the Institut d'Astrophysics de
Paris, whose kind hospitality I wish to acknowledge. I am grateful to
Stefano Borgani, Andrea Lapi, Gary Mamon, and Piero Rosati for helpful
comments and conversations.  I also wish to thank Alfonso Cavaliere
and Yoel Rephaeli for inviting me to give these lectures at the course
on the ``Astrophysics of Galaxy Clusters'' of the International School
of Physics ``Enrico Fermi''. Finally, it is my pleasure to dedicate
this work to my wife Patrizia, whose continuous support has proven
fundamental to the success of it.

\end{document}